\documentclass[aps,preprint,preprintnumbers,amsmath,amssymb,floatfix]{revtex4}
\usepackage{epsfig}
\usepackage{graphicx}
\usepackage{bm}
\usepackage{amssymb}
\usepackage{dcolumn}
\usepackage{bm}
\begin{document}
\draft

\title
{Motion-selective coherent population trapping for subrecoil cooling of optically trapped atoms outside the Lamb-Dicke regime}
\author{Hyun Gyung Lee, Sooyoung Park, Meung Ho Seo, and D. Cho\footnote{e-mail address:{\tt cho@korea.ac.kr}}}
\affiliation{Department of Physics, Korea University, Seoul 02841, Korea}

\date{\today}

\begin{abstract}
We propose a scheme that combines velocity-selective coherent population trapping (CPT) and 
Raman sideband cooling (RSC) for subrecoil cooling of optically trapped atoms outside the Lamb-Dicke regime.
This scheme is based on an inverted $\mathsf{Y}$ configuration in an alkali-metal atom.
It consists of a $\Lambda$ formed by two Raman transitions between the ground hyperfine levels
and the $D$ transition, allowing RSC along two paths and formation of a CPT dark state.
Using state-dependent difference in vibration frequency of the atom in a circularly polarized trap, 
we can tune the $\Lambda$ to make only the motional ground state a CPT dark state.
We call this scheme motion-selective coherent population trapping (MSCPT). 
We write the master equations for RSC and MSCPT
and  solve them numerically for a $^{87}$Rb atom in a one-dimensional optical lattice 
when the Lamb-Dicke parameter is 1.
Although MSCPT reaches the steady state slowly compared with RSC, the former consistently produces colder atoms than the latter.
The numerical results also show that subrecoil cooling by MSCPT outside the Lamb-Dicke regime is possible
under a favorable, yet experimentally feasible, condition.
We explain this performance quantitatively 
by calculating the relative darkness of each motional state.
Finally, we discuss on application of the MSCPT scheme to an optically trapped diatomic polar molecule
whose Stark shift and vibration frequency exhibit large variations depending on the rotational quantum number.

\end{abstract}

\maketitle

\section{INTRODUCTION}
Recoil by emission of final photons is the last hurdle in laser cooling atoms to a standstill.
Besides evaporative cooling, which entirely avoids laser lights, 
two schemes have been developed to overcome the hurdle:
Raman sideband cooling (RSC) \cite{Wineland1989}  for trapped atoms 
and velocity-selective coherent population trapping (VSCPT) \cite{VSCPT 1988} for free atoms.
Both methods achieve subrecoil cooling by using an arrangement that makes the motional ground state 
dark owing to either energy conservation or quantum interference,  respectively.
Although they are efficient tools, they are applicable in rather limited cases. 
RSC, which was originally developed for ions in a tight trap, can achieve subrecoil cooling
only when the vibrational energy spacing $\hbar \nu$ of a trap is much larger than the recoil energy $\mathcal{E}_R$,
or equivalently when the Lamb-Dicke parameter $\eta_{LD}$, defined by $\eta_{LD}^2 = \mathcal{E}_R/\hbar \nu$, 
is much less than 1. 
For optically trapped neutral atoms, the condition is not satisfied unless a lattice configuration with submicron confinement is employed. 
Subrecoil cooling by VSCPT has been demonstrated only for metastable He atoms with zero nuclear spin.
Efforts to apply the scheme to alkali-metal atoms, such as gray molasses \cite{gray molasses cooling}, 
have achieved only sub-Doppler cooling, 
and VSCPT is not applicable to trapped atoms.

In this paper, we propose a cooling method that combines VSCPT and RSC so that
they complement each other to overcome the limits they have when applied separately. 
Using the method, we aim to achieve subrecoil cooling of alkali-metal atoms in an optical trap 
even when the Lamb-Dicke condition is not satisfied. 
If we approach the aim starting from VSCPT, there are three main issues: 
(i) Owing to the hyperfine structure, any $\Lambda$ configuration formed by a pair of $D$ transitions of an alkali-metal atom has a leakage path out of it, complicating arrangement for 
coherent population trapping (CPT) in a steady state.
(ii) There is no velocity selection for bound-state atoms, 
 and we need a scheme that selects the motional ground state as a CPT dark state.
(iii) VSCPT by itself is only a diffusive process \cite{Levy flight}, 
and an extra cooling mechanism is needed, especially in 2D and 3D. 
(i) For the leakage problem, we have proposed an ``inverted $\mathsf{Y}$" configuration \cite{M1 CPT} 
consisting of a $\Lambda$ formed by two ground hyperfine transitions 
from the states $|\phi_1 \rangle$ and $|\phi_2 \rangle$ to the apex state $|\phi_3 \rangle$ 
that is coupled to the excited state $|\phi_4 \rangle$ by the $D$ transition (Fig. 1).
Using $^7$Li in an optical trap, we have demonstrated that the CPT phenomena of the inverted $\mathsf{Y}$ in a wide range of experimental parameters  could be precisely described by a leak-free $\Lambda$ system.
(ii) For the motional selectivity, we use a circularly polarized trap beam.
The vector polarizability $\beta$ causes 
$|\phi_1 \rangle$ and $|\phi_2 \rangle$ to have different well depths, and hence,
 different vibration frequencies $\nu_1$ and $\nu_2$, respectively,
as shown in Fig. 2.
Thus, two-photon detuning between the motional states 
$|\phi_1, \chi_1(n) \rangle$ and $|\phi_2, \chi_2(n) \rangle$ 
depends on the vibrational quantum number $n$,
and we can tune the $\Lambda$ fields so that only the $n=0$ pair forms a CPT dark state. 
We call this scheme motion-selective coherent population trapping (MSCPT). 
(iii) For the cooling, we propose to replace two radio frequency (rf) fields used in our previous work \cite{M1 CPT}
with two pairs of Raman beams, each of which is red detuned for the sideband cooling. 
From the viewpoint of RSC, 
by adding $|\phi_2 \rangle \rightarrow |\phi_3 \rangle$ transition 
to a usual one of $|\phi_1 \rangle \rightarrow |\phi_3 \rangle$, 
we have a $\Lambda$ configuration with a possibility of forming a CPT state.
The difference between $\nu_1$ and $\nu_2$  allows us to select the pair of $n=0$ states for CPT,
providing them an extra protection from the recoil heating.
Another advantage is that there is no need for repumping atoms fallen to the $|\phi_2 \rangle$ state, 
which reduces the average recoil heating for an optical pumping cycle in MSCPT.

\begin{figure}[t] \centering
	\includegraphics[scale=0.3]{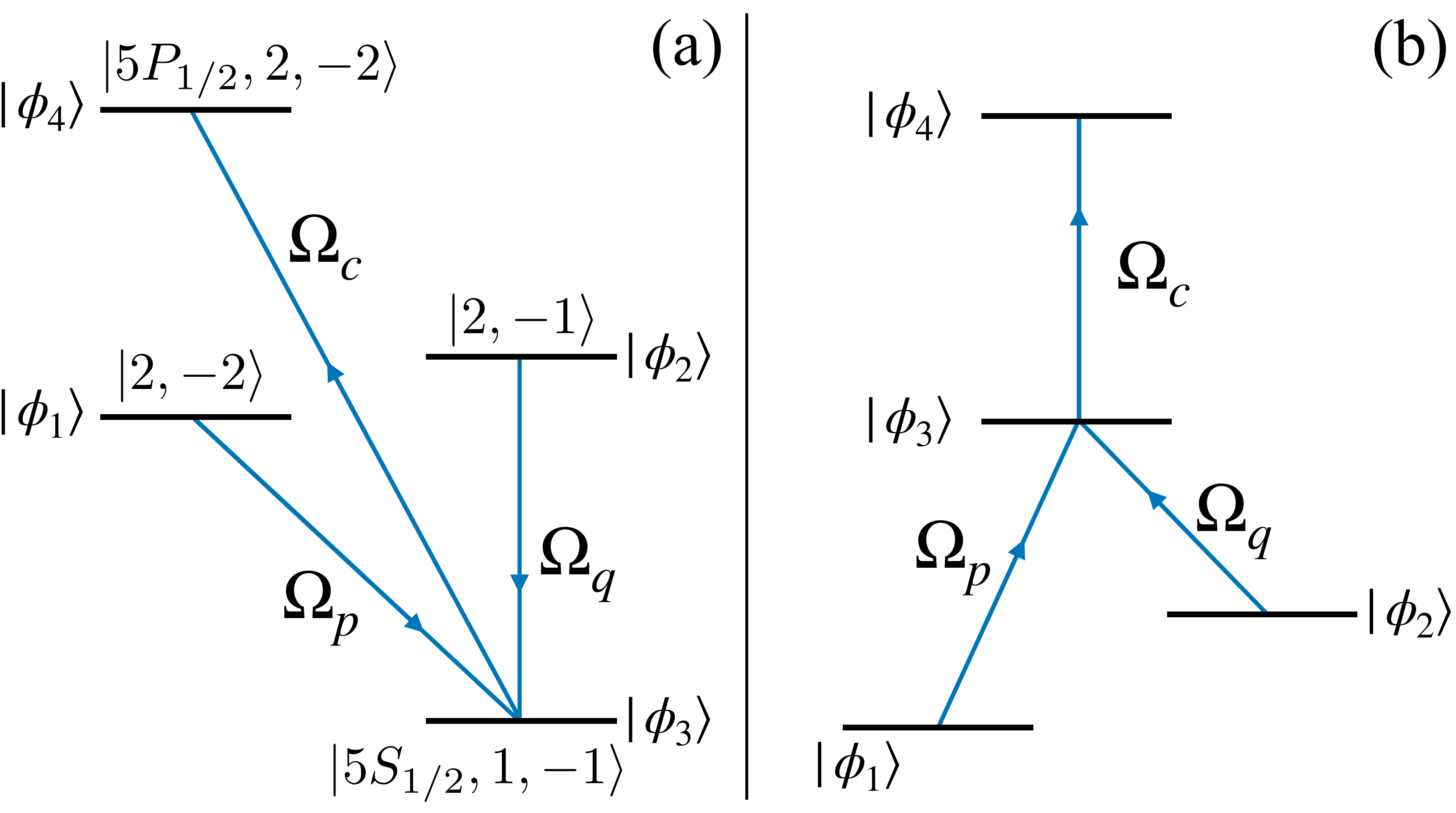}
	\caption {\baselineskip 3.0ex 
	(a) Raman transitions $\Omega_p, \Omega_q$ and the $D1$ transition $\Omega_c$ in $^{87}$Rb for the cooling scheme of motion-selective coherent population trapping. 
		(b) Inverted $\mathsf{Y}$ configuration for more intuitive visualization of MSCPT scheme.}
\end{figure}

\begin{figure}[t] \centering
	\includegraphics[scale=0.3]{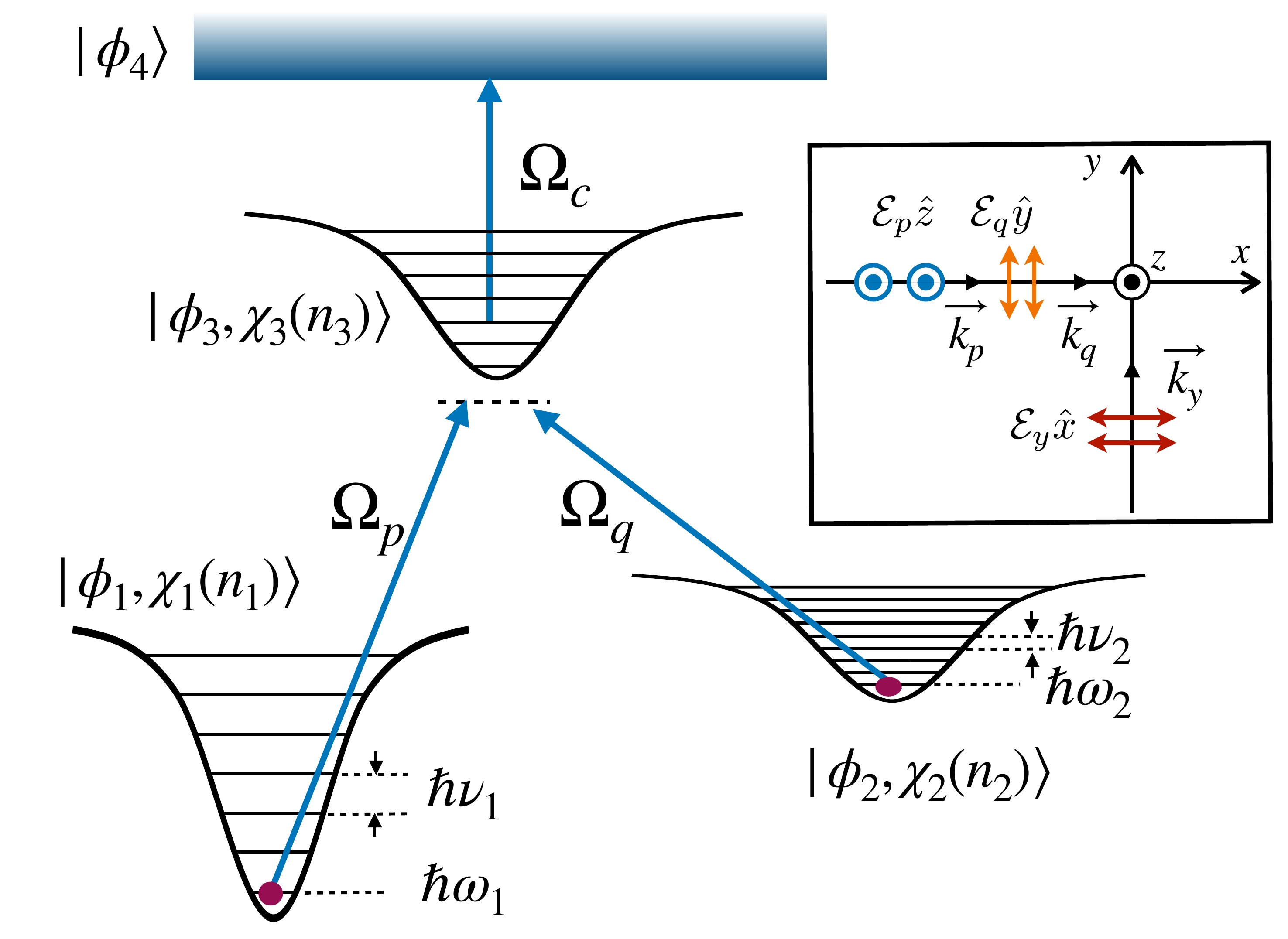}
	\caption[Magic polarization: Lineshape of a Rabi transition]{\label{magic_Rabi}\small   Inverted $\mathsf{Y}$ configuration for an atom in an optical trap. 
		$p$ and $q$ transitions are red-detuned for sideband cooling, and
		difference between the vibration frequencies $\nu_1$ and $\nu_2$ is responsible for motional selectivity. 
		The difference is exaggerated.
		$\hbar \omega_1$ and $\hbar \omega_2$ are the ground state energy of a trapped atom 
		in the $|\phi_1\rangle$ and $|\phi_2\rangle$ state, respectively.
		The inset shows an arrangement of three laser beams $\vec{E}_p, \vec{E}_q$, and $\vec{E}_y$ for $p$ and $q$ transitions. 
		Here, momentum transfer $\hbar\Delta\vec{k}_p$ and $\hbar\Delta \vec{k}_q$ to an atom by the two Raman transitions are the same.}
\end{figure}

There are two critical parameters for the success of the MSCPT cooling scheme: 
the difference $\Delta \nu_{12} = \nu_1 - \nu_2$ that brings the motional selectivity, 
and the coherence decay rate $\gamma_{12}$ that destroys it. 
Because $\Delta \nu_{12}$ is proportional to $\beta$, heavy alkali-metal atoms with large spin-orbit coupling are favored.  
For a $^{87}$Rb atom in an optical trap at wavelength 980 nm, $\Delta\nu_{12}$ is $2 \pi \times$ 25 Hz when $\eta_{LD} =1$,
whereas the full width at half maximum (FWHM) of the CPT resonance in the inverted $\mathsf{Y}$ was 150 Hz
in the rf experiment.
Sources that contribute to $\gamma_{12}$ are
fluctuations in magnetic field and a phase noise between the pair of Raman beams.
In Ref. \cite{M1 CPT}, we reduced $\gamma_{12}$ to 1.5 s$^{-1}$, which corresponds to FWHM of 0.25 Hz in rf spectroscopy, by shielding the ambient field and controlling the current noise. 
$\gamma_{12}$ from the Raman phase noise can be easily reduced below 1 s$^{-1}$ 
using modulation and phase-locking techniques.
Nevertheless, this implies that cooling by MSCPT requires precautions normally reserved for precision spectroscopy.
When the atomic density is high, collisions can dephase a CPT state \cite{collisional dephasing}, 
and the $D$ beam, that couples $|\phi_3 \rangle$ to $|\phi_4 \rangle$, 
may mediate photo-association and subsequent heating and loss of the atoms.
In this regard, MSCPT is best suited for cooling single atoms in an optical lattice or a tweezer. 
In addition, cooling by MSCPT is slower than that by RSC because even a pair with nonzero $n$ can form a partially dark superposition state, hampering the sideband cooling.
Some high-$n$ states may also form parasitic CPT dark states.

In spite of these difficulties, we recently demonstrated effectiveness of the MSCPT scheme in an experiment using $^{87}$Rb atoms in a 1D optical lattice \cite{MSCPT experiment}. 
We observed CPT phenomena driven by a pair of stimulated Raman transitions 
and, by employing the MSCPT scheme, achieved lower temperature than that obtained by RSC. 
Finally, we envision using MSCPT to cool optically trapped polar molecules \cite{polar molecule},
whose Stark shift exhibits strong dependence on the rotational quantum number.
However, for this application, finding an appropriate configuration for robust CPT is a prerequisite.

In the following sections, we describe the MSCPT scheme in more detail
and present the master equations and the results of the numerical simulations.
Using RSC as a benchmark, we evaluate performance of MSCPT 
in terms of the steady-state distribution of atoms over $n$ and dynamics toward it.

\section{Motion-selective coherent population trapping}
The backbone of the MSCPT scheme is the arrow-like configuration shown in Fig. 1 (a), 
where we use $^{87}$Rb as an example. 
In Fig. 1(b), it is transformed to an inverted $\mathsf{Y}$ configuration for more intuitive visualization.
It consists of a $\Lambda$ formed by the ground hyperfine transitions 
$|\phi_1\rangle = |5S_{1/2}, F= 2, m_F =-2 \rangle \rightarrow |\phi_3\rangle = |F= 1, m_F =-1 \rangle$ 
and $|\phi_2\rangle = |F= 2, m_F =-1 \rangle \rightarrow |\phi_3\rangle $,
which we will call $p$ and $q$ transitions, respectively,  
and the $D1$ coupling from the apex state $|\phi_3\rangle$
to the excited state, $|\phi_4\rangle =|5P_{1/2}, F= 2, m_F =-2 \rangle$.
Here, $F$ is the total angular momentum and $m_F$ is its $z$ component. 
The $D1$ coupling opens a path for $|\phi_3\rangle$ to decay to a CPT dark state via $|\phi_4\rangle$.
Angular momentum conservation dictates that 
$|\phi_4 \rangle$ decays only to one of the three states in the $\Lambda$,
and the inverted $\mathsf{Y}$ is closed. 
The $D1$ is favored over the $D2$ coupling owing to the simple hyperfine structure of the $5P_{1/2}$ state.
An equivalent inverted $\mathsf{Y}$ configuration can be identified in all alkali-metal atoms.
The master equations for the inverted $\mathsf{Y}$ \cite{master eq for inverted Y}
can be reduced to those for an effective $\Lambda$ system by adiabatically eliminating $|\phi_4 \rangle$ 
owing to its short lifetime \cite{M1 CPT}.
Unlike a typical $\Lambda$ formed by two $D$ couplings,
this effective $\Lambda$ allows us to choose the decay rate $R$ of the  $|\phi_3\rangle$ state by adjusting the $D1$  coupling strength as follows:
\begin{equation}
	R = \frac{1}{2} \frac{|\Omega_c|^2/2}{\Delta_c^2+\Gamma^2/4} \Gamma,
	\label{eq: R1}
\end{equation}
where $\Gamma$ is the decay rate of $|\phi_4\rangle$ 
and $\Omega_c$ and $\Delta_c$ are the Rabi frequency and the detuning of the $D1$ coupling, respectively.

Considering a $^{87}$Rb atom in an optical trap of a circularly polarized Gaussian beam 
with peak intensity $I_0$, 
the trap depth for the $|5S_{1/2}, F, m_F \rangle$ state is \cite{analogous Zeeman}	
\begin{equation}
U_0 (F, m_F) = (\alpha  + \beta g_F m_F ) (2\mu_0 c I_0),
\label{eq:U0}
\end{equation}
where $\alpha$ and $\beta$ are the scalar and vector polarizabilities, respectively, 
and $g_F$ is the Land\'{e} $g$ factor.
The fractional change in the vibration frequency of the $|F, m_F \rangle$ state
with respect to $\nu_0$ of the $m_F =0$ state is $(\beta/2\alpha) g_F m_F$.
If the Raman fields, denoted by $\Omega_p$ and $\Omega_q$ in Fig. 2, 
are tuned to the $\Lambda$ transition between the pair of motional ground states, 
the $|\phi_1, \chi_1(n) \rangle$ and $ |\phi_2, \chi_2(n) \rangle $ states 
have a two-photon detuning $ n\Delta\nu_{12}$, 
making their CPT-like superposition state progressively brighter as $n$ increases. Here, 
\begin{equation}
	\Delta\nu_{12} = \frac{\beta}{4\alpha}\,\nu_0
	\label{eq:Delta_nu_12}
\end{equation} 
for the configuration in Fig. 1(a).

In MSCPT, each of the $|\phi_1, \chi_1(n_1) \rangle$ and $|\phi_2, \chi_2(n_2) \rangle $ states can make stimulated Raman transitions to $|\phi_3, \chi_3(n_3) \rangle$ for a range of $n_3$. 
The Rabi frequency for the $p$ transition $|\phi_1,\chi_1(n_1) \rangle \rightarrow |\phi_3,\chi_3(n_3) \rangle$ is
\begin{equation}
	\Omega_p(n_3, n_1) =\Omega_p^0 \,\mathcal{F}_{31} (n_3, n_1),
	\label{eq:Omega_p}
\end{equation}
where $\Omega_p^0$ is the Rabi frequency for a free atom and the Franck-Condon factor is defined by  
\begin{equation}
	\mathcal{F}_{31} (n_3, n_1) =\langle \chi_3(n_3) |\,  e^{i\Delta \vec{k}_p\cdot \vec{r}}\, | \chi_1(n_1) \rangle.
	\label{eq:FCF}
\end{equation}
Here, $\hbar\Delta \vec{k}_p$ is the linear momentum transfer by the pair of Raman beams for the $p$ transition
and $\vec{r}$ is the center-of-mass coordinate of the atom. 
Similarly, for the $q$ transition $|\phi_2,\chi_2(n_2) \rangle \rightarrow |\phi_3,\chi_3(n_3) \rangle$,
$\Omega_q(n_3, n_2) =\Omega_q^0 \,\mathcal{F}_{32} (n_3, n_2) $,
where $\hbar \Delta \vec{k}_q$ is the momentum transfer. 
For the pair $\{ |\phi_1, \chi_1(n) \rangle, |\phi_2, \chi_2(n) \rangle \} $ to form a CPT dark state, 
the transition amplitudes for all allowed $p$ and $q$ transitions to the two respective groups of $|\phi_3, \chi_3(n_3) \rangle$ 
should interfere destructively. 
This is possible if $\Delta \vec{k}_p = \Delta \vec{k}_q$ so that 
$\mathcal{F}_{31} (n_3, n) =\mathcal{F}_{32} (n_3, n)$  and the pair of states share the target group.
Inset in Fig. 2 shows one arrangement of laser beams for the Raman transitions in Fig. 1 (a)
that satisfy $\Delta \vec{k}_p = \Delta \vec{k}_q$.
Specifically, the $p$ transition is driven by $\vec{E}_y= \mathcal{E}_y \hat{x} \cos(k_y y -\omega_y t)$ and $\vec{E}_p= \mathcal{E}_p \hat{z} \cos(k_p x -\omega_p t)$ 
with $\Delta \vec{k}_p = \vec{k}_p - \vec{k}_y$,
and the $q$ transition by  $\vec{E}_y$ and $\vec{E}_q= \mathcal{E}_q \hat{y} \cos(k_q x -\omega_q t)$ with $\Delta \vec{k}_q=\vec{k}_q-\vec{k}_y$.
Here, 
$\Delta \vec{k}_p = \Delta \vec{k}_q$ to a very good approximation  
with a discrepancy originating from the Zeeman shift of less than 1 MHz 
between $|\phi_1\rangle$ and $|\phi_2 \rangle $.
Because $|\chi_1(n) \rangle$ and $|\chi_2(n) \rangle$ are different, 
$\mathcal{F}_{31} (n_3, n)$ and $\mathcal{F}_{32} (n_3, n)$ are not identical, 
and this may complicate the CPT formation as well. 
However, this discrepancy affects, to first order, only the amplitudes of $|\phi_1, \chi_1(n) \rangle$ and $|\phi_2, \chi_2(n) \rangle$ in a dark superposition state, and the simulations show that the effects are insignificant 
for the experimentally feasible range of $\Delta \nu_{12} /\nu_0$.

\section{Master Equations}
Although the 2D configuration in the inset of Fig. 2 is a natural realization of the MSCPT scheme, 
we limit our discussion to the master equations and their numerical solutions in 1D. 
Extending the formalism to 2D is burdensome but straightforward.
However,  with our computing resources, numerical simulations are feasible only in 1D 
when states with a sufficiently large $n$ are included.
We also focus on single atoms, and ignore collisions between them.

\subsection{Raman sideband  cooling in 1D}
For 1D RSC, we consider a system consisting of only the states, 
$|\phi_1, \chi_1 (n_1) \rangle, \, |\phi_3, \chi_3 (n_3) \rangle$, and $ |\phi_4 \rangle$ in Fig. 2.
The Hamiltonian for this system is
\begin{equation}
	H_{\rm{RSC}}  = H_0 +W_p +V,
	\label{eq: H for RSC}
\end{equation}
where $H_0$ is for a trapped atom, $W_p$ for the $p$ transition, 
and $V$ for dissipative processes.
Although the difference between $\nu_1$ and $\nu_3$ is irrelevant in RSC, 
we distinguish $|\chi_1 \rangle$ and $|\chi_3 \rangle$ to ensure a formalism consistent with that of MSCPT.
The master equation for the density matrix $\rho$ is
\begin{equation}
	i\hbar \frac{d \rho}{dt} = [H_0 + W_p, \rho \,] +\left( \frac{\partial \rho}{\partial t} \right), 
	\label{eq: master eq}
\end{equation}
where the second term represents the dissipative processes. 
Explicitly, 
\begin{subequations}
	\begin{eqnarray}
H_0 &=& \sum_{j=1,3} \sum_{n_j^{\prime \prime}}
                        (\hbar \omega_j+\hbar \nu_jn_j^{\prime \prime} )
                        |\phi_j, \chi_j(n_j^{\prime \prime}) \rangle \langle \phi_j, \chi_j(n_j^{\prime \prime})|,
\label{eq: H0 for RSC} \\
W_p &=& \frac{\hbar \Omega_p^0}{2}  e^{-i \omega_p^{\,\prime} t} \sum_{n_1^{\prime \prime}, n_3^{\prime \prime}} 
                \mathcal{F}_{31}(n_3^{\prime \prime}, n_1^{\prime \prime}) 
                |\phi_3, \chi_3(n_3^{\prime \prime}) \rangle \langle \phi_1, \chi_1(n_1^{\prime \prime})|
                + \mbox{h.c.},
\label{eq: Wp for RSC}
	\end{eqnarray}
\end{subequations}
where $\hbar \omega_j$ is the energy of the lowest vibrational state $|\phi_j, \chi_j (0) \rangle$,
and $\omega_p^{\,\prime} = \omega_y - \omega_p$.
In the 1D formalism, $\Delta \vec{k}_p = k_p\hat{x}$ is substituted in Eq. (\ref{eq:FCF}); 
$\mathcal{F}_{31}(n_3, n_1) = \langle \chi_3(n_3) | \, e^{ i k_p x} \, |\chi_1(n_1) \rangle$.
The master equations, 
which include the decay of the $|\phi_3 \rangle$ state at the rate $R$ in Eq. (\ref{eq: R1}), 
can be expressed in terms of 
$\eta_{jj}(n_j, n_j^\prime) = \rho_{jj}(n_j, n_j^\prime)$ for $j=1,3$, and
$\eta_{13} (n_1, n_3) = \rho_{13}(n_1, n_3) e^{-i\omega_p^\prime t}$ as follows:

\begin{subequations}
	\begin{eqnarray}
		\label{eq:RSC OBE1}
\dot{\eta}_{11}(n_1, n_1^\prime) 
&=& 
  i\frac{\Omega_p^0}{2}  \sum_{n_3^{\prime\prime}} 
   \left\{\eta_{13}(n_1,n_3^{\prime\prime}) \mathcal{F}_{31} (n_3^{\prime\prime},n_1^\prime)
   - \mathcal{F}_{13}^{\,*} (n_1, n_3^{\prime\prime}) \eta_{31}(n_3^{\prime\prime}, n_1^\prime)\right\} \\ \nonumber 
 &+& i (n_1^\prime -n_1) \nu_1\eta_{11}(n_1, n_1^\prime)
       +  \delta_{n_1, n_1^\prime} \, 
       p_1 R  \sum_{n_3^{\prime\prime}} |\mathcal{F}_{13} (n_1, n_3^{\prime\prime})|^2
       \eta_{33}(n_3^{\prime \prime}, n_3^{\prime \prime}),  \\ 
       \label{eq:RSC OBE2} 
\dot{\eta}_{33}(n_3, n_3^\prime)
&=& 
i\frac{\Omega_p^0}{2}  \sum_{n_1^{\prime\prime}} 
\left\{\eta_{31}(n_3, n_1^{\prime\prime}) \mathcal{F}_{13}^{\,*} (n_1^{\prime\prime},n_3^\prime)
- \mathcal{F}_{31} (n_3, n_1^{\prime\prime}) \eta_{13}(n_1^{\prime\prime}, n_3^\prime)\right\} \\ \nonumber
&+& \{i (n_3^\prime -n_3) \nu_3-R\}\eta_{33}(n_3, n_3^\prime)
+\delta_{n_3, n_3^\prime} \,
p_3 R  \sum_{n_3^{\prime\prime}} |\mathcal{F}_{33} (n_3, n_3^{\prime\prime})|^2
\eta_{33}(n_3^{\prime \prime}, n_3^{\prime \prime})  ,  \\  
     \label{eq:RSC OBE3}
\dot{\eta}_{13}(n_1, n_3)  	
&=& 
i\frac{\Omega_p^0}{2}\{ \sum_{n_1^{\prime\prime}} 
\eta_{11}(n_1, n_1^{\prime\prime}) \mathcal{F}_{13}^{\,*} (n_1^{\prime\prime},n_3)
- \sum_{n_3^{\prime\prime}} \mathcal{F}_{13}^{\,*} (n_1, n_3^{\prime\prime}) \eta_{33}(n_3^{\prime\prime}, n_3)\} \\ \nonumber
&+& [i\{ (n_3+\Delta n)\nu_3 -n_1 \nu_1\}-R/2]\eta_{13}(n_1, n_3),
	\end{eqnarray}
\end{subequations}
and $\dot{\eta}_{31}(n_3, n_1)  	=\dot{\eta}_{13}^{\,*} (n_1, n_3)$.
We use the rotating wave approximation.
$\Delta n$, which is defined as $\omega_p^{\,\prime} -(\omega_3 - \omega_1)  = -\Delta n \nu_3$, 
represents the order of the red sideband. 
The last terms in Eqs. (\ref{eq:RSC OBE1}) and (\ref{eq:RSC OBE2}) describe
the decay of $|\phi_3, \chi_3 (n_3^{\prime \prime}) \rangle$
to $|\phi_1, \chi_1 (n_1) \rangle$ and $|\phi_3, \chi_3 (n_3) \rangle$
with branching ratios $p_1$ and $p_3$, respectively,   
as a process of emitting a photon with a momentum $\hbar k_p$.
The $\mathcal{F}$ factors satisfy the relation \cite{Cohen Tanouudij},
\begin{equation}
	\sum_{n_i^{\prime \prime}=0}^{\infty} 
	E_i(n_i^{\prime \prime}) |\mathcal{F}_{ij}(n_i^{\prime \prime}, n_j)|^2 -E_j(n_j) =\mathcal{E}_R,
	\label{eq: sum rule}
\end{equation}
and the recoil heating by $\mathcal{E}_R = \hbar^2 k_p^2/2m$, where $m$ is the atomic mass, 
accompanying the decay is built into the master equations. 
Here, $E_j(n_j) = \hbar\nu_j (n_j+1/2), j = 1,3,$ is the motional energy of the $|\chi_j(n_j) \rangle$ state.
However, the real process is an excitation to $|\phi_4 \rangle$ by an absorption 
and a subsequent decay by a spontaneous emission. 
When averaged over the angular distribution of the emission, the total heating is by $2\mathcal{E}_R$.
In numerical simulations, we take this into account 
by using the branching ratios $p_1^\prime = p_1/2$  and $p_3^\prime = 1-p_1/2$ 
to double the number of emissions required to optically pump an atom from $|\phi_3 \rangle$ to $|\phi_1 \rangle$. 

In an experiment, parametric heating from trap noise
and depumping of the $|\phi_1 \rangle$ state by an imperfect optical pumping
are common problems.
The transition rate from the motional state $|\chi(n) \rangle$ to $|\chi(n \pm 2) \rangle$ driven 
by the intensity noise of a trap beam is approximated as
\begin{equation}
	Q_{\pm}(n) = \frac{\pi \nu^2}{16} S(2\nu) (n+1 \pm1)(n\pm1),
	\label{eq: PHR}
\end{equation}
where $S(2\nu)$ is the power spectral density of the fractional intensity noise at twice the vibration frequency \cite{parametric heating}.
Its effect on the time evolution of the population $P_j(n) = \eta_{jj}(n,n)$ and the coherence ${\eta}_{ij}(n_i, n_j)$  can be included by adding the following new terms:
\begin{subequations}
	\begin{eqnarray}
		\label{eq:PHR_pop}
		\dot{P}_j(n)
		&\Leftarrow& 
		 -Q (n)P_j(n) +Q_+ (n-2)P_j(n-2)+Q_- (n+2)P_j(n+2),
		\\
	\label{eq:PHR_coh}
		\dot{\eta}_{ij}(n_i, n_j)  	
		&\Leftarrow& 
		-\frac{1}{2} \{Q(n_i)+Q( n_j)\} \,\eta_{ij}(n_i,n_j), 
	\end{eqnarray}
\end{subequations}
where $Q(n) = Q_+(n)+Q_-(n)$. 
If the effective decay rate $D$ of the $|\phi_1\rangle$ state caused by its unintended transition 
to an excited state $|\phi_e \rangle$ is included,
\begin{subequations}
	\begin{eqnarray}
		\label{eq:OPL_pop}
		\dot{P}_j(n)
		&\Leftarrow& 
		-DP_j(n) \delta_{j1} +p_j D \sum_{n_1^{\prime\prime}}
		 |\mathcal{F}_{j1} (n_j, n_1^{\prime\prime})|^2 P_1(n_1^{\prime\prime}),
		\\
	\label{eq:OPL_coh}
		\dot{\eta}_{ij}(n_i, n_j)  	
		&\Leftarrow& 
		-\frac{1}{2}D (\delta_{i1} +\delta_{j1}) \,\eta_{ij}(n_i,n_j), 
	\end{eqnarray}
\end{subequations}
where we assume that $|\phi_e \rangle$ has the same branching ratios $p_1$ and $p_3$ as $|\phi_4 \rangle$.

\subsection{Motion-selective coherent population trapping in 1D}
For 1D MSCPT in Fig. 2, $H_{\rm{MSCPT}}=H_{\rm{RSC}}+W_q$ and $H_0$ for a trapped atom in Eq. (\ref{eq: H0 for RSC}) is augmented by
\begin{equation}
\sum_{n_2^{\prime \prime}=0}^{\infty}
(\hbar \omega_2+\hbar \nu_2 n_2^{\prime \prime})
|\phi_2, \chi_2(n_2^{\prime \prime}) \rangle \langle \phi_2, \chi_2(n_2^{\prime \prime})|.
\label{eq: H0 for MSCPT}
\end{equation}
$W_q$ for the $q$ transition is
\begin{equation}
W_q = \frac{\hbar \Omega_q^0}{2}  e^{-i \omega_q^{\,\prime} t} \sum_{n_2^{\prime \prime}, n_3^{\prime \prime}} 
\mathcal{F}_{32}(n_3^{\prime \prime}, n_2^{\prime \prime}) 
|\phi_3, \chi_3(n_3^{\prime \prime}) \rangle \langle \phi_2, \chi_2(n_2^{\prime \prime})|
+ \mbox{h.c.},
\label{eq: Wq for MSCPT}
\end{equation}
where $\omega_q^{\,\prime} = \omega_y - \omega_q$.
The angular momentum selection rule and the condition $|\,\omega_1 -\omega_2\,| \gg R $ forbid 
the $p$ Raman fields from driving the $q$ transition, and vice versa. 
The master equations for 1D MSCPT are listed in Appendix A.
Below, we write the equation for $\dot{\eta}_{12}(n_1, n_2)$ only 
because it includes the critical terms that describe the motional selectivity and the decay of the CPT coherence.
\begin{eqnarray}
	\dot{\eta}_{12}(n_1, n_2)
	&=&  	\frac{i}{2}\sum_{n_3^{\prime\prime}} 
 \left\{ \Omega_q^0 \eta_{13}(n_1, n_3^{\prime\prime}) \mathcal{F}_{32} (n_3^{\prime\prime},n_2)
-  \Omega_p^0 \mathcal{F}_{13}^{\,*} (n_1, n_3^{\prime\prime}) \eta_{32}(n_3^{\prime\prime}, n_2)\right\}  
\label{eq:eta_12} \\
   &+&
\{ i (n_2\nu_2-n_1\nu_1-\delta_{\rm{CPT}})-\gamma_{12} \} \eta_{12}(n_1, n_2),	\nonumber
\end{eqnarray}
where $\delta_{\rm{CPT}} = (\omega_p^{\, \prime} -\omega_q^{\, \prime}) -(\omega_2 -\omega_1)$ 
is the detuning of the $p$ and $q$ Raman fields from the CPT resonance of the motional ground states
and  $\gamma_{12}$ is the coherence decay rate.
The effects of parametric heating and depumping can be incorporated as done in RSC.

\section{Numerical Simulations}
Given experimental parameters for either RSC or MSCPT, 
we are interested in the distribution $P(n) =\sum_j \eta_{jj}(n,n)$ 
of atoms in a steady state and dynamics toward it.
By reshaping $\eta_{ij}(n_i,n_j)$ into an $N$-dimensional column vector $x$, 
the master equations can be written as
$\dot{x} = Ax$ with an appropriately defined matrix $A$.
If the maximum $n$ to be included in a calculation is $n_c$, 
$N = 4(n_c+1)^2$ for 1D RSC and $9(n_c+1)^2$ for 1D MSCPT.
A steady-state solution $x_s$ satisfies $Ax_s=0$ under the constraint $\sum_n P(n) =1$.
We use the Moor-Penrose algorithm for pseudo-inversion to solve the equation.
When $n_c = 100$, the run time on a personal computer to obtain $x_s$ is 3 min for RSC and 30 min for MSCPT.
Time evolution of $x$ follows $x(t) = e^{At}x(0)$.
We calculate $U(\tau) =e^{A\tau}$ for an appropriate time interval $\tau$ 
and obtain snap shots of $x(t)$ at $t = \tau, 2\tau, ..$ by repeatedly applying $U(\tau)$.
The calculation of $U(\tau)$ is demanding in terms of time and memory, 
and the run time for RSC with $n_c = 75$ is 1 h.
For MSCPT, we limit $n_c$ to  65 and use single precision to calculate $U(\tau)$ in 2 h.
For the numerical solutions, 
we need the values of the Franck-Condon factors $\mathcal{F}_{ij} (n_i, n_j)$.
When $\eta_{LD} \geq 1$, for a given $n_j$, the range of $n_i$ to be calculated is large, 
and the evaluation of each $\mathcal{F}_{ij} (n_i, n_j)$ is time-consuming 
because the polynomial expansion of $e^{ik_p x}$ does not converge. 
To efficiently calculate the  $\mathcal{F}$ factors,  we develop recursion relations. 
The relations are summarized in Appendix B. 
The calculated results are validated by the sum rule $\sum_{n_i^{\prime \prime}=0}^{\infty} |\mathcal{F}_{ij}(n_i^{\prime \prime}, n_j)|^2 =1$, based on the completeness of  $|\chi_i (n_i) \rangle$ and the unitarity of $e^{i k_p x}$.
In Appendix B, we also include the case of $k_p =0$ for a transition by an rf field.

As a model system, we use a 1D optical lattice in our apparatus \cite{RF spectroscopy}.
Its wavelength $\lambda_{\rm OL}$ is 980 nm, 
at which $\alpha =-873$ and $\beta = -25$ in atomic units  for $^{87}$Rb. 
We adjust the minimum spot size $w_0$ to 10 $\mu$m 
and the well depth in units of the Boltzmann constant $k_B$ to 125 $\mu$K 
so that $\nu_0 = 2\pi \times 3.5$ kHz and $\eta_{LD}\simeq 1$ for the transverse motion.
When the lattice beam is circularly polarized, $\Delta \nu_{12}$ is $2\pi \times 25$ Hz.
For a $^{87}$Rb atom in Fig. 1(a), the branching ratios from $|\phi_4 \rangle$ 
to $|\phi_1 \rangle, |\phi_2\rangle$, and $|\phi_3 \rangle$ are 
$p_1=1/3$, $p_2=1/6$, and $p_3 =1/2$, respectively. 
On average, three optical pumping cycles are needed to put an atom into $|\phi_1 \rangle$ in RSC, 
and two cycles to put it into either $|\phi_1 \rangle$ or $|\phi_2 \rangle$ in MSCPT.
However, we use $p_1 =1/2$ for RSC and $p_1 = p_2 =1/4$ for MSCPT in the following simulations
to compare the cooling efficiencies while the recoil heating rates are the same. 
In addition, as discussed previously,
we use $p_1^\prime  =1/4$ for RSC and $p_1^\prime=p_2^\prime =1/8$ for MSCPT
to take into account additional heating from the absorption of an optical pumping photon.
We use $R, \Omega_p$, and $\Omega_q$ equal to $\nu_0$, and 
$\gamma_{12} = 3 \times 10^{-4} \nu_0$ or $ 2\pi \times 1$ Hz as benchmark values.
The two pairs of Raman beams are tuned to $\delta_{\rm CPT} =0$
so that the $n=0$ states are CPT resonant. 

\begin{figure}[t] \centering
	\includegraphics[scale=0.2]{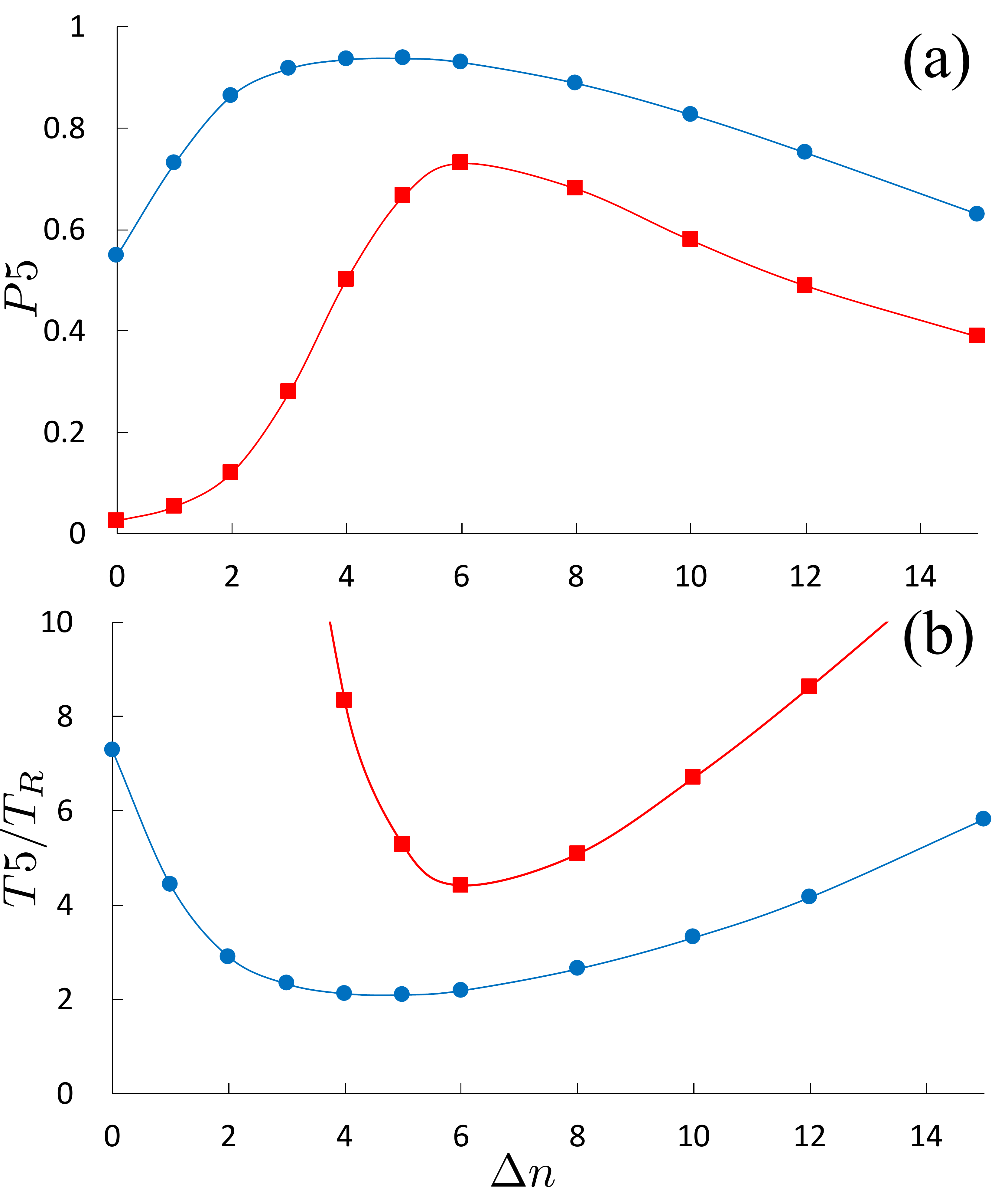}
	\caption[Magic polarization: Lineshape of a Rabi transition]{\label{magic_Rabi}\small    (a) $P5$, sum of the steady-state population with the motional quantum number $n$ below 5,
		versus the order $\Delta n$ of the red sideband.
		We use the benchmark condition described in the main text.
		Red squares are for RSC, and blue circles are for MSCPT.
		(b) $T5$, temperature calculated from $P5$ assuming the Maxwell-Boltzmann distribution, 
		in units of $T_R=\mathcal{E}_R/k_B$  versus  $\Delta n$. 
		$\mathcal{E}_R$ is the recoil energy accompanying emission of a photon
		and $k_B$ is the Boltzmann constant.
		The lowest $T5$ produced by MSCPT is $2T_R$, while that by RSC is $4.5T_R$.}
\end{figure}

First, we calculate $P(n)$ versus the order $\Delta n$ of the red sideband for both RSC and MSCPT.
Figure 3(a) shows $P5 = \sum_{n=0}^5 P(n)$ versus $\Delta n$,
and Fig. 3(b) shows $T5 = -6\hbar\nu_0/k_B \ln(1-P5)$ in units of $T_R = \mathcal{E}_R/k_B$.
$T_R= 175$ nK, and
$T5$ is the temperature that produces $P5$ under the Maxwell-Boltzmann (MB) distribution.
$T5$ is a better measure of temperature than one by fitting $P(n)$ to the MB distribution
because without atomic collisions, 
$P(n)$ reflects the details of the $\mathcal{F}$ factors and does not follow the MB distribution.
We choose $P5$ because $dP5/dT$ is maximum at around $T=3T_R$, 
the temperature range of interest in Fig. 3(b).
The minimum $\Delta n$ to overcome the recoil heating in 
either RSC or MSCPT is 4 when $\eta_{LD} = 1$.
For RSC, $\Delta n =6$ produces the lowest $T5$
owing to the radiative broadening of the Raman transition by $R$. 
Reducing $R$ results in the optimal order $\Delta n_{\rm opt}$ approaching 4 and a lower temperature;
however, this is at the expense of slower cooling. 
Although the simulation produces smooth reduction of $P5$ to finite values when $\Delta n \leq 3$, 
it is an artifact of truncating $n$ at $n_c$.
In an experiment, atoms are expected to boil out in this condition. 
For MSCPT, $\Delta n_{\rm opt}=5$ and $P5$ shows a gradual decrease away from it.
MSCPT consistently produces a lower temperature than RSC, 
with the lowest $T5$ of $2T_R$ compared with $4.5T_R$ by RSC.
If the real value of $p_1 = 1/3$ for RSC is used, $\Delta n_{\rm opt}$ is 7, 
and the minimum $T5$ is $6T_R$, three times higher than that by MSCPT. 
\begin{figure}[b] \centering
	\includegraphics[scale=0.2]{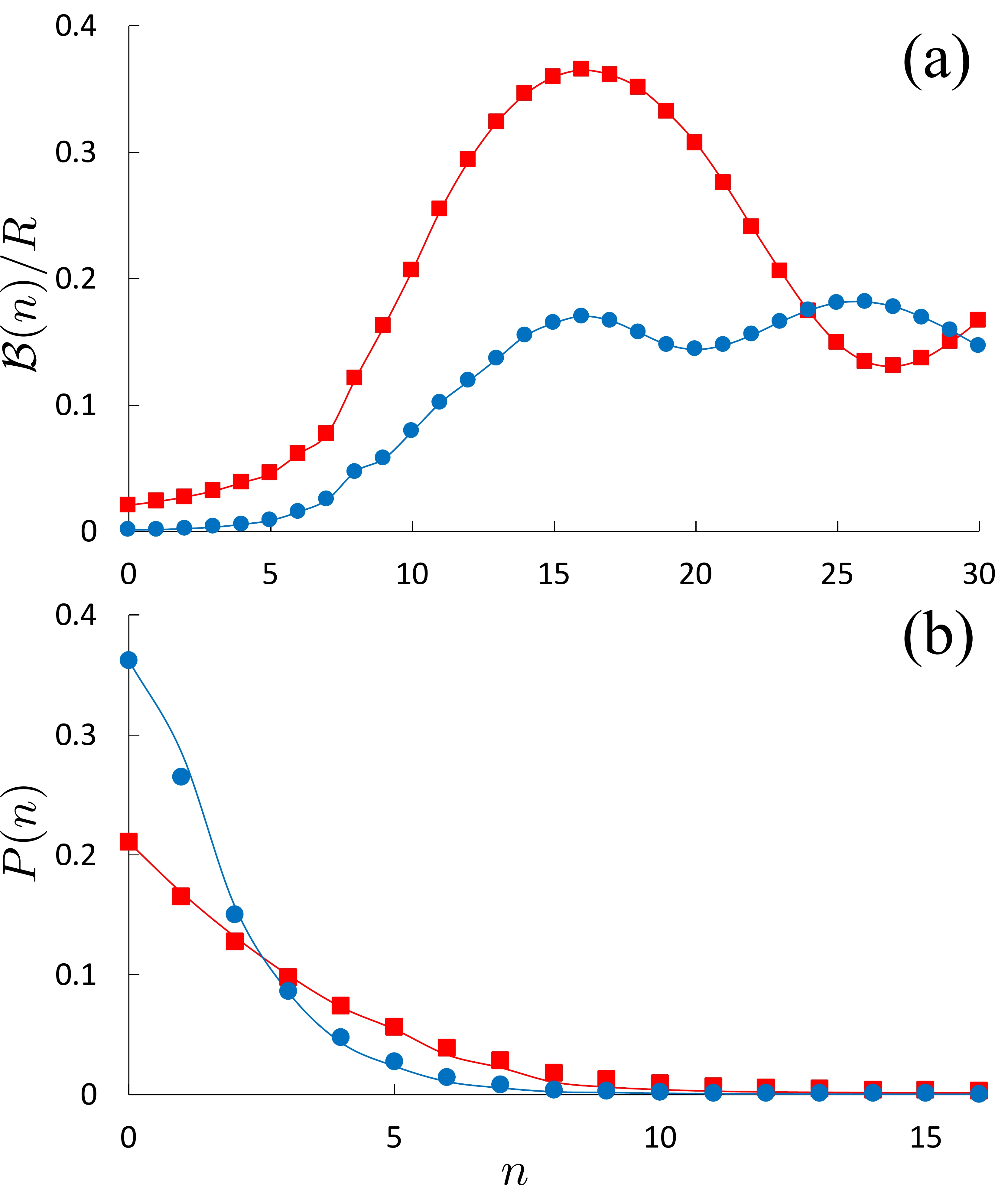}
	\caption[Magic polarization: Lineshape of a Rabi transition]{\label{magic_Rabi}\small  Comparison of RSC and MSCPT in terms of the brightness $\mathcal{B}(n)$ of the $n$th state. $\mathcal{B}(n)$ is defined as a product of the effective decay rate $R$ of the $|\phi_3 \rangle$ state and the total population in the $|\phi_3 \rangle$ state.
		(a) $\mathcal{B}(n)$ for RSC (red square) and MSCPT (blue circle) in units of $R$.
		The low-$n$ states in MSCPT are much darker than those in RSC.	
		(b) The steady-state population $P(n)$ versus $n$. 
		$P(n)$ is inversely proportional to $\mathcal{B}(n)$. 
		$P(n)$ is fitted by the red curve $P(0)[\mathcal{B}(0)/\mathcal{B}(n)]^{1.7}$ for RSC
		and by the blue curve $P(0)[\mathcal{B}(0)/\mathcal{B}(n)]^{1.35}$ for MSCPT.}
\end{figure}

The better performance of MSCPT is a consequence of the CPT-induced darkness of the low-$n$ states. 
We define the brightness of the $n$th pair of states $\{ |\phi_1, \chi_1(n) \rangle, |\phi_2, \chi_2(n) \rangle\}$ in MSCPT 
as a product of $R$ and the population in the $|\phi_3 \rangle$ state,
 \begin{equation}
 \mathcal{B}(n) = R \sum_{n_3 = 0}^{n_c} P_3(n_3).
 \label{eq: Brightness}
 \end{equation} 
$\mathcal{B}(n)$ of the $n$th state $|\phi_1, \chi_1(n) \rangle$ in RSC can be similarly defined. 
We obtain $\mathcal{B}(n)$  for MSCPT or RSC by solving the master equations in a reduced Hilbert space 
consisting of the $n$th pair or the $n$th state, respectively, and $\{|\phi_3, \chi_3(n_3) \rangle, n_3 = 0, 1, ..n_c\}$.
Figure 4(a) shows $\mathcal{B}(n)/R$ at the respective $\Delta n_{\rm opt}$ of RSC and MSCPT. 
The low-$n$ states in MSCPT are significantly darker than those in RSC.
We expect $P(n)$ to be inversely proportional to $\mathcal{B}(n)$.
In Fig. 4(b), the red and blue curves of $P(0) [ \mathcal{B}(0)/\mathcal{B}(n) ]^a$ with $a =1.7$ and 1.35
show good agreement with $P(n)$ of RSC and MSCPT, respectively.  
One drawback of this darkness is the slowdown of the cooling process. 
Figure 5 shows the evolution of $T5$ under the optimal RSC and MSCPT starting from the MB distribution at 3 $\mu$K.
Both curves follow a double exponential decay expressed as
\begin{equation}
T5(t) = (T_i - T_m) e^{-t/\tau_1} +(T_m - T_f) e^{-t/\tau_2} +T_f
\end{equation}
because, unlike a simple decay of the same entities,
a qualitative change occurs in the atomic ensemble as the cooling proceeds. 
Here, $T_i, T_m,$ and $T_f$ are the initial, middle, and final temperatures, respectively, 
and $\tau_1$ and $\tau_2$ are time constants.
Although MSCPT produces a lower $T_f$, it is 5 times slower than RSC; specifically, 
$\tau_1 = 1.1$ ms and $\tau_2 = 7.7$ ms for RSC and 
$\tau_1 = 5.2$ ms and $\tau_2 = 34$ ms for MSCPT.
\begin{figure}[t] \centering
	\includegraphics[scale=0.27]{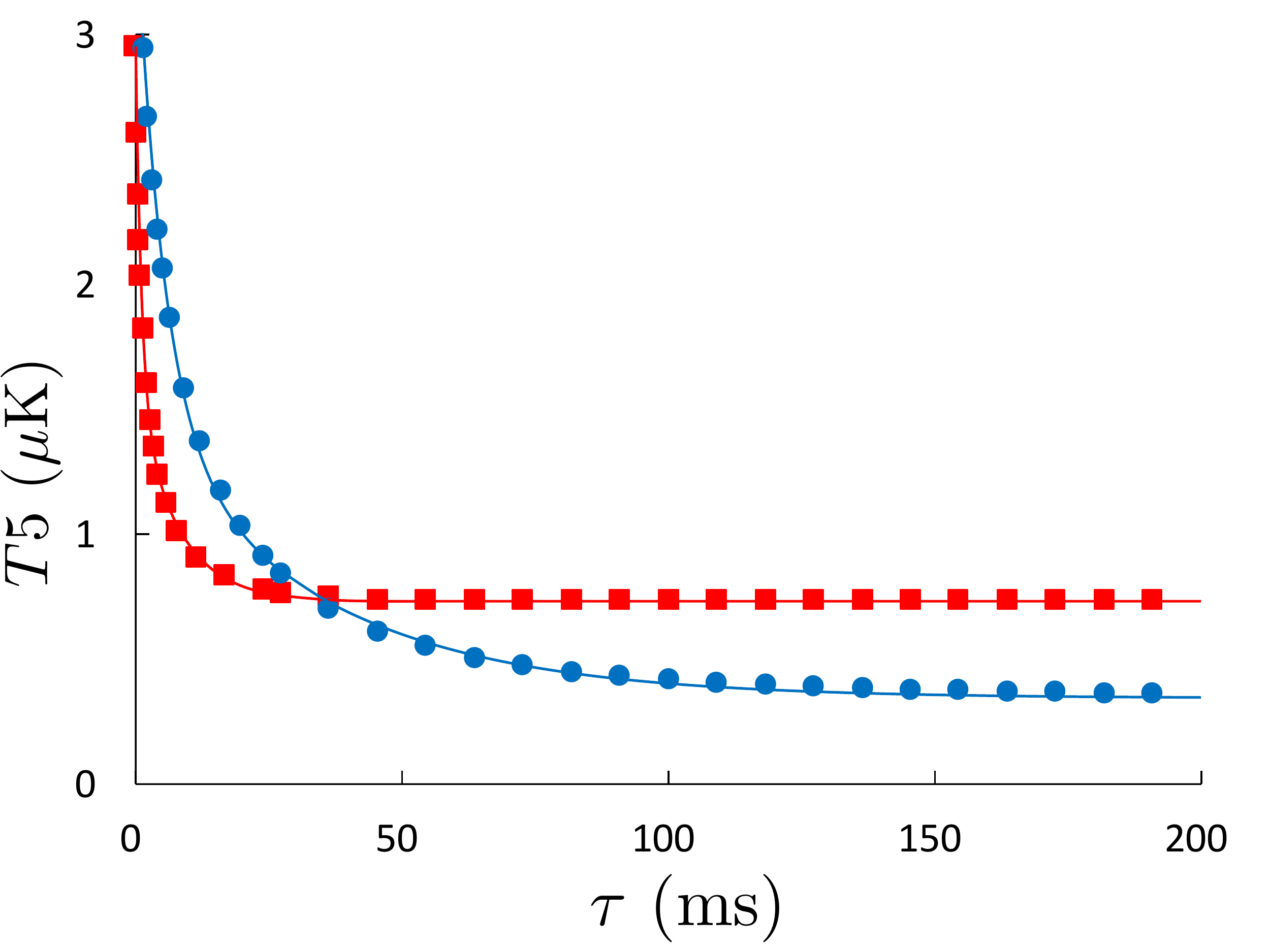}
	\caption[Magic polarization: Lineshape of a Rabi transition]{\label{magic_Rabi}\small Evolution of $T5$ for RSC (red square) and MSCPT (blue circle) 
		starting from the MB distribution at 3 $\mu$K.
		They are fitted by double exponential decay curves with 
		$\tau_1 = 1.1$ ms and $\tau_2 = 7.7$ ms for RSC (red solid line), and
		$\tau_1 = 5.2$ ms and $\tau_2 = 34$ ms for MSCPT (blue solid line).}
\end{figure}

Next, we change $\delta_{\rm CPT}$ of MSCPT while keeping $\Delta n=5$.
$T5/T_R$ versus $\delta_{\rm CPT}$ in units of $\Delta \nu_{12}$ is shown in Fig. 6(a).
The dashed horizontal line denotes $T5/T_R$ for RSC.
According to Eq. (\ref{eq:eta_12}) for $\dot{\eta}_{12}(n_1, n_2)$, 
the $n$th pair of states $\{ |\phi_1, \chi_1(n) \rangle, |\phi_2, \chi_2(n) \rangle\}$ 
is CPT resonant when $\delta_{\rm CPT} = -n \Delta\nu_{12}$. 
As $\delta_{\rm CPT}$ becomes negative,  $T5$ increases sharply as 
the$n = 1, 2, ..$ pairs successively become dark.
The minimum $T5$ of RSC, $4.5 T_R$, corresponds to the average $n$ of 4,
to which $T5$ of MSCPT becomes comparable when $\delta_{\rm CPT} =-5\Delta\nu_{12}$.
The inset of Fig. 6(a) shows $P(n)$ peaks at $n=5$ when $\delta_{\rm CPT} = -5\Delta\nu_{12}$.
For a positive $\delta_{\rm CPT}$, although detuning from the CPT resonance increases for all $n$ pairs, 
that of the $n=0$ pair is still the smallest.
This explains the gradual increase in $T5$ for $\delta_{\rm CPT} >0$. 
Figure 6(b) shows $P(0)$ versus $\delta_{\rm CPT}/\Delta\nu_{12}$. 
$P(0)$ is maximum not at $\delta_{\rm CPT}=0$ but at $1.3\Delta\nu_{12}$
because $d\mathcal{B}(n)/dn$ or the contrast is more critical than $\mathcal{B}(n)$ itself for determining $P(n)$,
and the detuning puts $\mathcal{B}(n)$ on a slope near $n=0$.
The inset shows $\mathcal{B}(n)$ for 
$\delta_{\rm CPT} = 0$ (black square) and $1.3\Delta\nu_{12}$ (red circle).
The width between the arrows in Fig. 6(a), where MSCPT shows noticeable advantage, 
is $10 \Delta\nu_{12}$ or 250 Hz.
While the Raman fields can be easily tuned to within 1 Hz or better, 
$d\delta_{\rm CPT}/dB = 350$ Hz/mG, where $B$ is the quantization field strength, 
and precise control of the B field is a more demanding task in practice. 

\begin{figure}[b] \centering
	\includegraphics[scale=0.22]{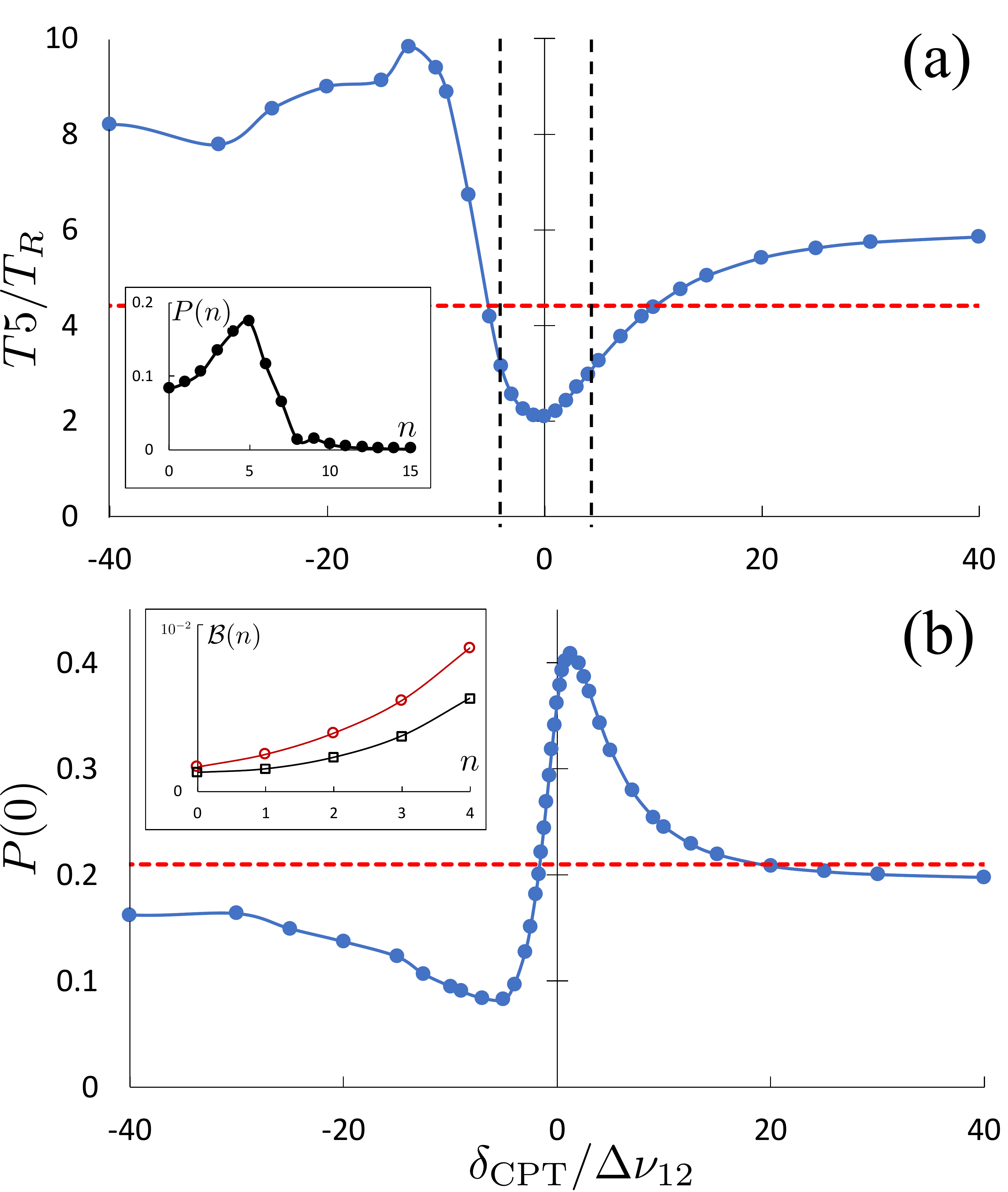}
	\caption[Magic polarization: Lineshape of a Rabi transition]{\label{magic_Rabi}\small    (a) $T5$ in units of $T_R$ versus $\delta_{\rm CPT}$ in units of $\Delta\nu_{12}$. 
		$\delta_{\rm{CPT}}$	is the detuning of the $p$ and $q$ Raman fields from the CPT resonance of the motional ground states.
		The dashed red horizontal line represents $T5/T_R$ for RSC.
		The inset shows $P(n)$ when $\delta_{\rm CPT} = -5\Delta\nu_{12}$.
		(b) The ground population $P(0)$ versus $\delta_{\rm CPT}/\Delta\nu_{12}$. 
		The inset shows that $\mathcal{B}(n)$ when $\delta_{\rm CPT} = 1.3\Delta\nu_{12}$ (red circle) 
		increases more rapidly than $\mathcal{B}(n)$ when $\delta_{\rm CPT} = 0$ (black square).
		This explains the shift of $P(0)$ maximum to $\delta_{\rm CPT}=1.3\Delta\nu_{12}$.}
\end{figure}

Effects of $R, \Omega_p$, and $\Omega_q$ on the steady-state solution and the time constants are similar,
and we put $R =\Omega_p = \Omega_q$, considering them as a single parameter.
In  Fig. 7(a), we plot $T5/T_R$ of RSC and MSCPT versus $R$. 
Each $T5$ is obtained at $\Delta n_{\rm opt}$ for a given $R$, and 
the dependence of $\Delta n_{\rm opt}$ on $R$ for RSC (red square) and MSCPT (blue circle)
is shown in the inset.
As $R$ becomes large, the Raman transition broadens 
to increase $\mathcal{B}(n)$ of the  low-$n$ states, 
and $T5$ of RSC increases at $d T5/dR = 0.5$ $\mu$K/$\nu_0$.
In comparison, in MSCPT, the darkness of the low-$n$ states is further protected by CPT, 
and $T5$ is almost constant, indicating its robustness as a cooling method. 
However, when $R$ is much smaller than $\nu_0$, RSC produces a lower $T5$ than MSCPT.
Here, the narrow Raman width causes a step-like change in $\mathcal{B}(n)$ 
across $n = \Delta n$ in RSC, whereas in MSCPT, 
the finite width of the CPT resonance tends to blur the contrast.
In practice, as Fig. 7(b) shows, 
$\tau_2$ for both RSC and MSCPT sharply increases as $R$ becomes smaller.
Although not shown in the figure, at $R = \nu_0/10$, $\tau_2$ reaches 0.75 s and 3 s for RSC and MSCP,
respectively, making it impractical to excessively reduce $R$.
In the opposite limit of a large $R$, the time constants of RSC become vanishingly small, 
whereas those of MSCPT do not significantly change because the cooling dynamics is limited by 
the diffusive process of population trapping.
\begin{figure}[b] 
	\includegraphics[scale=0.2]{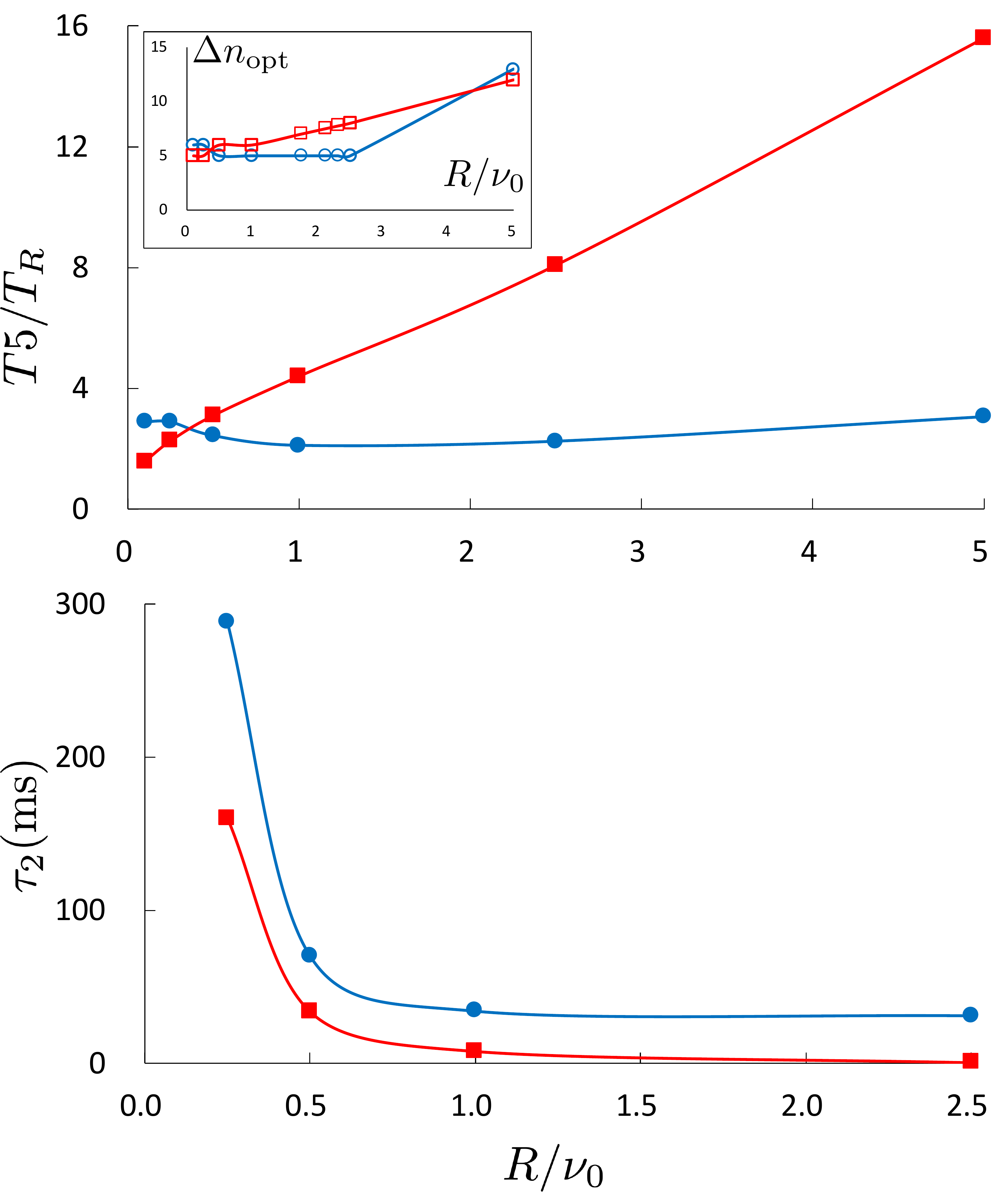}
	\caption[Magic polarization: Lineshape of a Rabi transition]{\label{magic_Rabi}\small (a) $T5/T_R$ of RSC (red square) and MSCPT (blue circle) 
		versus $R =\Omega_p = \Omega_q$ in units of $\nu_0$.
		Each $T5$ is obtained at an optimal $\Delta n$ for a given $R$. 
		The dependence of the optimal $\Delta n$ on $R$ is shown in the inset.
		(b) Time constant $\tau_2$ of RSC and MSCPT versus $R =\Omega_p = \Omega_q$ in units of $\nu_0$.}
\end{figure}

\begin{figure}[t] 
	\includegraphics[scale=0.2]{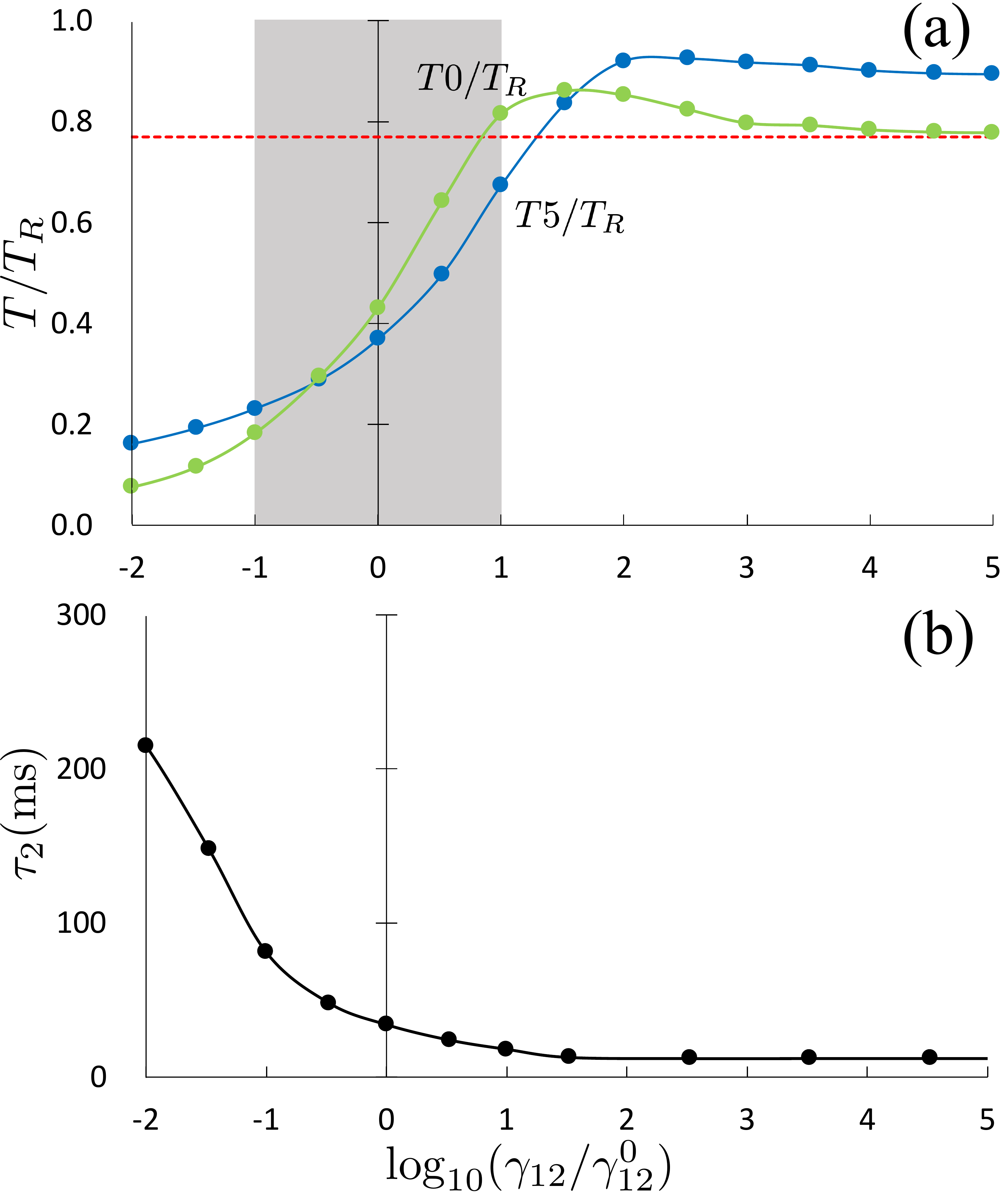}
	\caption[Magic polarization: Lineshape of a Rabi transition]{\label{magic_Rabi}\small  (a) $T5/T_R$ (blue circle) and $T0/T_R$ (green circle) of MSCPT versus $\log_{10}(\gamma_{12}/\gamma_{12}^{0})$.  
		$\gamma_{12}^{0} = 2\pi \times 1$ Hz, and $T0$ is defined from the ground population $P(0)$.
		The red dashed line represents for $T5/T_R$  of RSC.
		The shaded area is where MSCPT is both effective and experimentally feasible.
		(b) The time constant $\tau_2$ of $T5$ versus $\log_{10}(\gamma_{12}/\gamma_{12}^{0})$.  
		Although the temperature is lower at small $\gamma_{12}$, the cooling process becomes slower. }
\end{figure}
The decoherence rate $\gamma_{12}$ is one of the most critical parameters in the MSCPT scheme.
$T5/T_R$ of MSCPT versus $\log_{10}(\gamma_{12}/\gamma_{12}^{0})$, with  $\gamma_{12}^{0} = 2\pi \times 1$ Hz, are shown as blue circles in Fig. 8(a).
The red dashed line represents $T5/T_R$  of RSC.
When the temperature is below $T_R$, $T5$ based on $P5$ is no longer a sensitive measure of temperature,
and in Fig. 8(a), we include $T0$ defined from the ground population $P(0)$ 
using $T0 = -\hbar\nu_0/k_B \ln(1-P(0))$ as green circles.
The discrepancy between $T5$ and $T0$ is a signature of the deviation from the MB distribution.
In the limit of small a $\gamma_{12}$, the CPT phenomenon becomes prominent, 
and the atoms accumulate in the $n=0$ state.
When $\gamma_{12} = \gamma_{12}^0/10$, $T0$ is as low as $T_R$.
It implies that, using MSCPT, subrecoil cooling is possible even when $\eta_{LD}=1$.
However, reducing $\gamma_{12}$ to $\gamma_{12}^0/10$ is challenging.
In our previous work on $^7$Li using rf fields \cite{M1 CPT}, we achieved $\gamma_{12}^0/4$, limited by the magnetic field noise.
The final limit on $\gamma_{12}$ for an atom in an optical lattice originates from 
the scattering rate $R_{\rm OL}$ of the lattice photons \cite{dephasing_photon_scattering}.
In our model system of $\lambda_{\rm OL}= 980$ nm, $w_0 = 10$ $\mu$m, and well depth of 125 $\mu$K,
 $R_{\rm OL}= \gamma_{12}^0/6$.
Another difficulty associated with a small $\gamma_{12}$ is 
the increase in $\tau_2$ of $T5$, as shown in Fig. 8(b).
It is difficult to calculate $\tau_2$ of $T0$ at a small $\gamma_{12}$ owing to the slow convergence, 
and we only estimate that it is a few times larger than $\tau_2$ of $T5$.
At the other limit of $\gamma_{12}$ larger than $30 \gamma_{12}^{0}$, 
Fig. 8(a) shows that MSCPT produces higher $T5$ than RSC.
The mechanism for this is unclear, and we suspect 
that the large $\gamma_{12}$ broadens the CPT width, reducing the contrast in the darkness. 
Considering these results,
the experimentally interesting range is 0.1 Hz $\le \gamma_{12}/2\pi \le$ 10 Hz, which is shaded in Fig. 8(a).

 \begin{figure}[b] 
 	\includegraphics[scale=0.2]{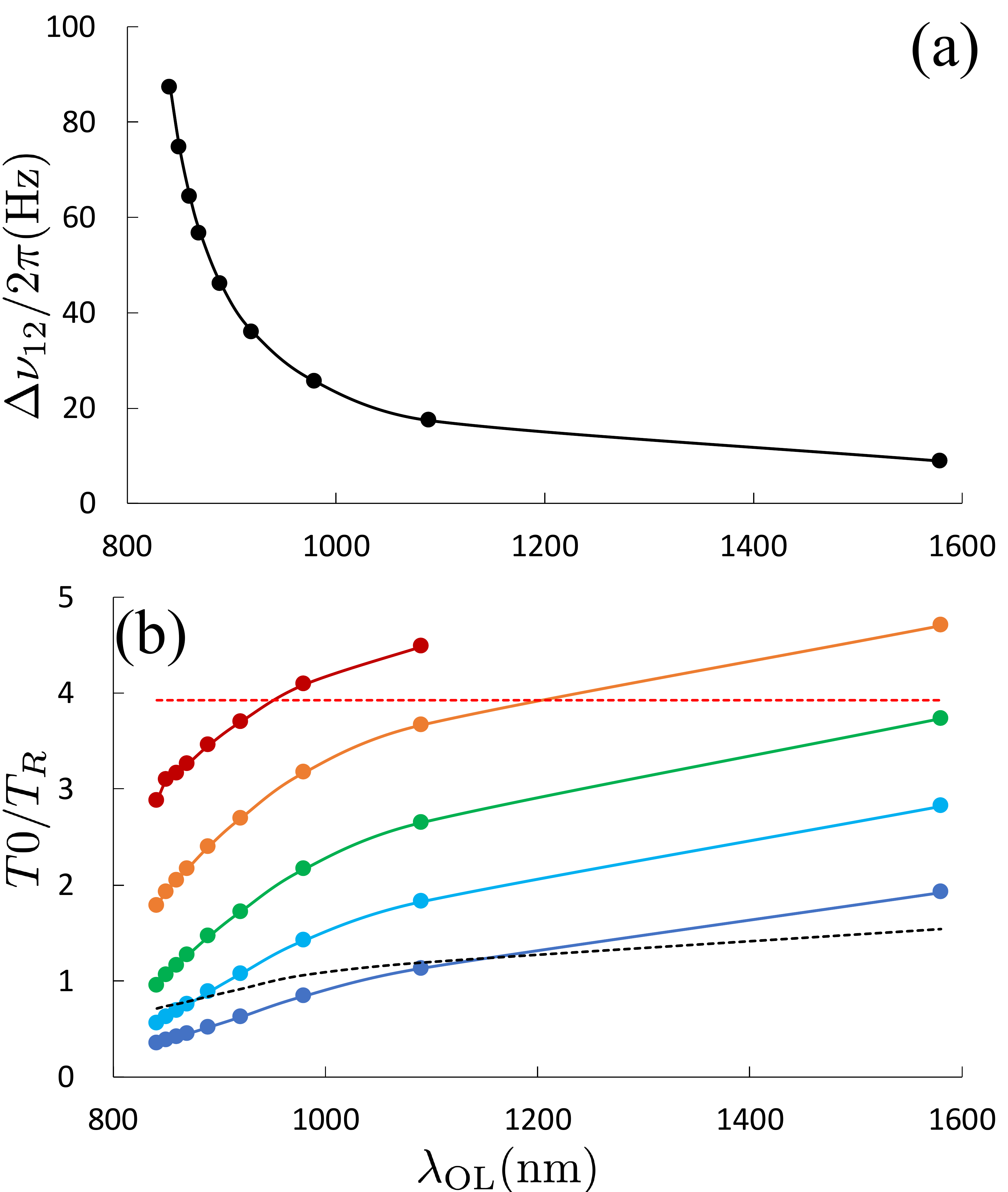}
 	\caption[Magic polarization: Lineshape of a Rabi transition]{\label{magic_Rabi}\small   (a) $\Delta \nu_{12}$ versus the lattice wavelength $\lambda_{\rm OL}$ 
 		when $w_0 = 10$ $\mu$m and the well depth in units of $k_B$ is 125 $\mu$K.
 		As $\lambda_{\rm OL}$ approaches the $D1$ transition, 
 		both $\Delta \nu_{12}$ and $\gamma_{12}$ increase as $1/\Delta_{D1}$.
 		$\Delta_{D1}$ is the detuning of $\lambda_{\rm OL}$ from the $D1$ transition.
 		(b) $T0/T_R$ of MSCPT versus $\lambda_{\rm OL}$  for a few $\gamma_{12}$.
 		From the uppermost to the lowest curve, $\gamma_{12}/2\pi$ = 10, 3, 1, 0.3, and 0.1 Hz. 
 		The red dashed line represents $T0/T_R$ of RSC.
 		The black dashed curve denotes $T0/T_R$ of MSCPT when $\gamma_{12} = R_{\rm OL}$ of a given $\lambda_{\rm OL}$.
 		The points below the curve are unattainable in our model system.}
 \end{figure}
We can also improve the motional selectivity by increasing $\Delta \nu_{12}$.
When $\lambda_{\rm OL}$ approaches 795 nm of the $D1$ transition, 
the ratio $\beta / \alpha$ and thereby $\Delta \nu_{12}$ at a fixed well depth increase
as $1/\Delta_{D1}$.
$\Delta_{D1}$ is the detuning of the lattice beam from the $D1$ transition.
Figure 9(a) shows $\Delta \nu_{12}$ versus $\lambda_{\rm OL}$ 
when $w_0 = 10$ $\mu$m and the well depth in units of $k_B$ is 125 $\mu$K so that $\eta_{LD}=1$.
At $\lambda_{\rm OL} = 841$ nm, $\Delta \nu_{12}$ is as large as $2\pi \times 87$ Hz,
facilitating the selection of the $n = 0$ state by CPT.
However, as $\lambda_{\rm OL}$ approaches the $D1$ transition,
the scattering rate $R_{\rm OL}$ at a fixed well depth
and consequently $\gamma_{12}$ also increase as $1/\Delta_{D1}$, offsetting the advantage.
Figure 9(b) shows $T0/T_R$ of MSCPT versus $\lambda_{\rm OL}$ 
for a few $\gamma_{12}/2\pi$ from 10 Hz of the uppermost curve to 0.1 Hz of the lowest one. 
The red dashed line represents $T0/T_R$ of RSC, and the black dashed curve shows $T0/T_R$ of MSCPT 
when $\gamma_{12} = R_{\rm OL}$ for a given $\lambda_{\rm OL}$.
The points below the curve are unattainable experimentally in our model system.

\begin{figure}[b] 
	\includegraphics[scale=0.2]{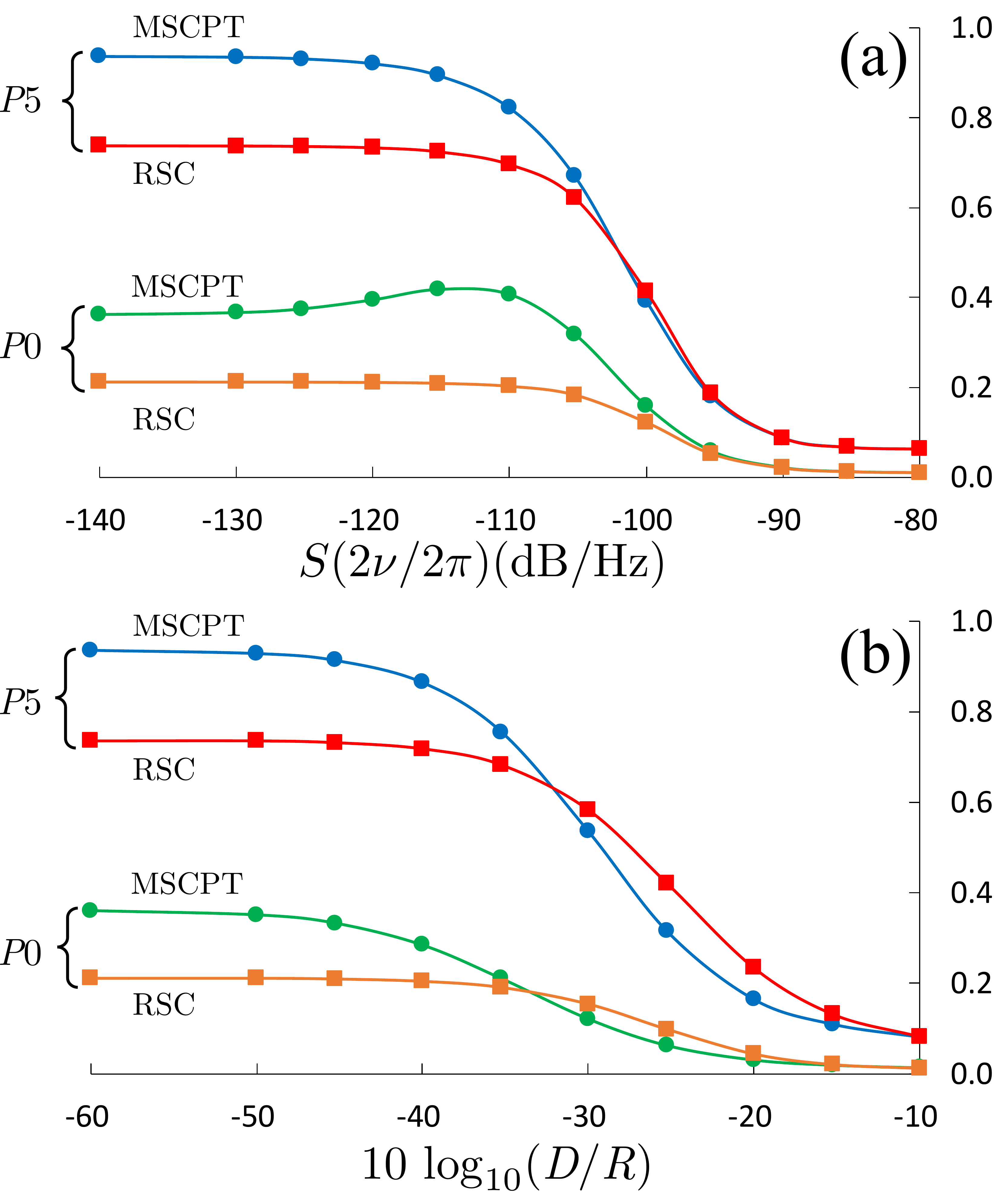}
	\caption[Magic polarization: Lineshape of a Rabi transition]{\label{magic_Rabi}\small (a) $P5$ of MSCPT (blue circle) and RSC (red square), 
		and $P0$ of MSCPT (green circle) and RSC (orange square) 
		versus $S(2\nu/2\pi)$, the power spectral density of the fractional intensity noise,  in dB/Hz.
		$P0$ of MSCPT has a peak at $S \simeq  - 115$ dB/Hz. 
		(b) $P5$ of MSCPT (blue circle) and RSC (red square), 
		and $P0$ of MSCPT (green circle) and RSC (orange square) versus $10\log_{10}(D/R)$. 
		$D$ is the effective decay rate of the target states, $|\phi_1 \rangle$ and $|\phi_2 \rangle$,
		due to their unintended transition by the optical pumping beam, 
		and the decay rate $R$ of the $|\phi_3 \rangle$ state is equal to $\nu_0$.	}
\end{figure}

Finally, we consider the effects of parametric heating and an imperfect optical pumping. 
They are respectively parametrized by the transition rate $Q_{\pm}(n)$ 
from $|\chi(n) \rangle$ to $|\chi(n \pm 2) \rangle$ in Eq. (\ref{eq: PHR}) and
the effective decay rate $D$ of the $|\phi_1\rangle$ and $|\phi_2\rangle$ states.
In Fig. 10(a), $P5$ and $P0$ of MSCPT and RSC versus 
the power spectral density of the fractional intensity noise $S(2\nu/2\pi)$ in dB/Hz are plotted.
$P5$ of MSCPT and RSC suffer similarly from the parametric heating, and 
its effect becomes insignificant when $S$ is below -125 dB/Hz. 
Noise can be lowered to this level by appropriate power stabilization  \cite{power_servo}.
$P0$ of MSCPT presents a peak at $S \simeq  - 115$ dB/Hz. 
The incoherent transition driven by the power noise contributes an additional dephasing rate $Q(n)$ to 
$\gamma_{12}$ of $\dot{\eta}_{12}(n,n)$
as expressed in Eq. (\ref{eq:PHR_coh}).
Because $Q_{\pm}(n)$ is approximately proportional to $n^2$,
relative darkness of the $n =0$ state is increased. 
This is similar to cooling by selective parametric excitation and subsequent ejection 
of high-$n$ states in an anharmonic optical trap \cite{cooling by anharmonicity}. 
In Fig. 10(b), $P5$ and $P0$ of MSCPT and RSC versus $10\log_{10}(D/R)$ are plotted, 
where the optical pumping rate $R$ is equal to $\nu_0$.
When the rate $D$ of the unintended transition out of the target states is five orders of magnitude 
smaller than the intended rate $R$,  
the effect of the imperfect optical pumping is negligible. 
$D$ may originate from an incorrect retardation $\delta \Gamma$ of a waveplate or
misalignment $\delta \theta$ between the quantization axis and the pumping beam direction.
$D/R$ is typically on the order $\delta \Gamma^2$ or $\delta \theta^2$, 
and these errors need to be kept below  a few times $10^{-3}$ radian.
The effect of $D$ on MSCPT is more severe than that on RSC 
because $D$ also degrades the coherence of a CPT state.
However, while RSC scheme on an alkali metal atom always requires repumping, 
the inverted $\mathsf{Y}$ configuration of MSCPT does not need it, 
and the target states $|\phi_1\rangle$ and $|\phi_2 \rangle$ are 6.8 GHz detuned from the optical pumping 
transition.  
We expect that, in practice, $D$ can be maintained much smaller in MSCPT.

\section{Discussion and Summary}
The 2D configuration of MSCPT in the inset of Fig. 2 does not provide cooling along the $z$ axis.
We may remedy the situation by applying another Raman beam
$\vec{E}_z= \mathcal{E}_z \hat{x} \cos(k_z z -\omega_z t)$
so that $(\vec{E}_z, \vec{E}_p)$ and $(\vec{E}_z,\vec{E}_q$) drive the $p$ and $q$ transitions, respectively.
Here, by adjusting $\omega_z$, the transitions are tuned to the red sideband of the $z$ motion. 
Apparently, the resulting double $\Lambda$ configuration increases the complexity of the scheme. 
However, in our model system of a 1D lattice, 
the vibration frequencies along the $z$ and $x,y$ axes differ by a factor of 50, and
once an atom falls to the $n_z =0$ state, it decouples from $\vec{E}_z$
because the Raman transitions are detuned by $50\nu_0$, 
while their widths are order $ \nu_0$. 
$n_z$ is the motional quantum number along $z$.  

The configuration in Fig. 1(a) is closed, and any of the three $5S_{1/2}$ states 
can play the role of the apex state $|\phi_3\rangle$ in Fig. 1(b). 
One interesting possibility is to exchange the roles of 
the $|F=1, m_F =-1\rangle$ and $|F=2, m_F =-1\rangle$ states in Fig. 1(a)
so that the latter becomes $|\phi_3\rangle$ in Fig. 1(b). 
In this configuration, $\Delta\nu_{12} =(3\alpha/4\beta)\nu_0$ is 
three times larger than the original $\Delta\nu_{12}$ in Eq. (\ref{eq:Delta_nu_12}), enhancing the motional selectivity.
However, whether this enhancement would lead to lower temperature is not clear because $\gamma_{12}$ from  magnetic-field noise is proportional to $\Delta\nu_{12}^2$.
$\gamma_{12}$ in this configuration is nine times larger, and 
unless noise of the field-generating current is tightly controlled, 
the gain in $\Delta \nu_{12}$ may be lost.

In summary, we propose a cooling scheme that combines the ideas of 
velocity-selective coherent population trapping and Raman sideband cooling.
Using the master equations for 1D RSC and 1D MSCPT, we calculate 
the steady-state distribution over $n$ and the time constants toward it
when the Lamb-Dicke parameter is 1,
as we change the experimental parameters such as
(i) the order of the sideband and the CPT detuning,
(ii) the optical pumping rate and the Rabi frequencies of the Raman transitions,
(iii)  the wavelength, and hence, the vector polarizability of an Rb atom, 
and the decoherence rate,
and (iv) the parametric heating rate and the depumping rate.
For most ranges of these parameters, 
MSCPT produces colder atoms than RSC, 
even though the recoil heating in RSC is reduced by two thirds by adjusting the branching ratio. 
Under a favorable condition, the temperature estimated from the $n=0$ population
reaches the recoil temperature $\mathcal{E}_R/k_B$ or below, 
indicating subrecoil cooling even outside the Lamb-Dicke regime. 
This improvement in cooling is quantitatively explained in terms of the  reduction in the brightness 
of the low-$n$ states by the CPT phenomenon.

However, this improvement has consequences. 
First, the MSCPT scheme is not suitable for a high-density atomic sample.
The optical pumping beam, which is near resonant and stays on throughout the cooling process,
mediates photo-association and atomic collisions destroy the coherence. 
The MSCPT scheme is best for a single atom in a lattice site or an optical tweezer.
Other problems are the slow cooling process and technical burdens to implement complex cooling beams 
and to control the phase noise from various sources.
Under these challenges,  a proof-of-principle experiment was successfully carried out in our laboratory to demonstrate the feasibility and effectiveness of the idea \cite{MSCPT experiment}, 
and an improved version of the experiment is  in progress.
Longer term, we plan to apply the MSCPT scheme to cool a diatomic polar molecule in an optical trap.
Using a MgF molecule as an example, $\Delta \nu_{12}$  of  the 
$|\phi_1 \rangle =|F =2, m_F=-2 \rangle$ and $|\phi_2 \rangle =|F =2, m_F=-1 \rangle$ states 
is 12\% of the average of $\nu_1$ and $\nu_2$ when the trap wavelength is 532 nm. 
The values are when the molecule is in the electronic and vibrational ground state and the first excited rotational state and its total electronic angular momentum is 3/2 and the nuclear spin is 1/2.
The X$^{2}\Sigma_{1/2}^{+}$ $\rightarrow$ A$^{2}\Pi_{1/2}$ transition of MgF is at 359 nm. 
In comparison, for $^{87}$Rb, $\Delta\nu_{12}/\nu_0$ is 0.7\% at the benchmark $\lambda_{\rm OL}$ = 980 nm and 2.5\% at 841 nm.
However, it is still unambiguous whether the complicated level structure of the molecule will allow
an appropriate configuration for the MSCPT scheme. 

\section*{ACKNOWLEDGMENTS}
This work was supported by the National Research Foundation of Korea (Grant No. 2019M3E4A1080382).
We thank Q-Han Park for help with the numerical work, D. G. Lee for assisting in the computer setup, 
and Eunmi Chae for discussions and information on molecules.

\newpage
\appendix
\section{Master equations for 1D motion-selective coherent population trapping}
For the Hamiltonian $H_{\rm{MSCPT}} = H_0+W_p+W_q+V$, 
where $H_0$ is the sum of Eqs. (\ref{eq: H0 for RSC}) and (\ref{eq: H0 for MSCPT}), 
$W_p$ and $W_q$ are given in Eq. (\ref{eq: Wp for RSC}) and Eq. (\ref{eq: Wq for MSCPT}), respectively,
and $V$ represents the radiative decay at $R$,
the master equations are
\begin{subequations}
	\begin{eqnarray}
	\label{eq:MSCPT OBE11}
	\dot{\eta}_{11}(n_1, n_1^\prime) 
	&=& 
	i\frac{\Omega_p^0}{2}  \sum_{n_3^{\prime\prime}} 
	\left\{\eta_{13}(n_1,n_3^{\prime\prime}) \mathcal{F}_{31} (n_3^{\prime\prime},n_1^\prime)
	- \mathcal{F}_{13}^{\,*} (n_1, n_3^{\prime\prime}) \eta_{31}(n_3^{\prime\prime}, n_1^\prime)\right\} \\ \nonumber 
	&+& i (n_1^\prime -n_1) \nu_1\eta_{11}(n_1, n_1^\prime)
	+  \delta_{n_1, n_1^\prime} \, 
	p_1 R  \sum_{n_3^{\prime\prime}} |\mathcal{F}_{13} (n_1, n_3^{\prime\prime})|^2
	\eta_{33}(n_3^{\prime \prime}, n_3^{\prime \prime}),  \\ 
	\label{eq: MSCPT OBE22}
	\dot{\eta}_{22}(n_2, n_2^\prime) 
	&=& 
	i\frac{\Omega_q^0}{2}  \sum_{n_3^{\prime\prime}} 
	\left\{\eta_{23}(n_2,n_3^{\prime\prime}) \mathcal{F}_{32} (n_3^{\prime\prime},n_2^\prime)
	- \mathcal{F}_{23}^{\,*} (n_2, n_3^{\prime\prime}) \eta_{32}(n_3^{\prime\prime}, n_2^\prime)\right\} \\ \nonumber 
	&+& i (n_2^\prime -n_2) \nu_2\eta_{22}(n_2, n_2^\prime)
	+  \delta_{n_2, n_2^\prime} \, 
	p_2 R  \sum_{n_3^{\prime\prime}} |\mathcal{F}_{23} (n_2, n_3^{\prime\prime})|^2
	\eta_{33}(n_3^{\prime \prime}, n_3^{\prime \prime}),  \\ 
	\label{eq:MSCPT OBE33} 
	\dot{\eta}_{33}(n_3, n_3^\prime)
	&=& 
	i\frac{\Omega_p^0}{2}  \sum_{n_1^{\prime\prime}} 
	\left\{\eta_{31}(n_3, n_1^{\prime\prime}) \mathcal{F}_{13}^{\,*} (n_1^{\prime\prime},n_3^\prime)
	- \mathcal{F}_{31} (n_3, n_1^{\prime\prime}) \eta_{13}(n_1^{\prime\prime}, n_3^\prime)\right\} \\ \nonumber
	&+& i\frac{\Omega_q^0}{2}  \sum_{n_2^{\prime\prime}} 
	\left\{\eta_{32}(n_3, n_2^{\prime\prime}) \mathcal{F}_{23}^{\,*} (n_2^{\prime\prime},n_3^\prime)
	- \mathcal{F}_{32} (n_3, n_2^{\prime\prime}) \eta_{23}(n_2^{\prime\prime}, n_3^\prime)\right\} \\ \nonumber
	&+& \{i (n_3^\prime -n_3) \nu_3-R\}\eta_{33}(n_3, n_3^\prime)
	+\delta_{n_3, n_3^\prime} \,
	p_3 R  \sum_{n_3^{\prime\prime}} |\mathcal{F}_{33} (n_3, n_3^{\prime\prime})|^2
	\eta_{33}(n_3^{\prime \prime}, n_3^{\prime \prime})  ,  \\  
	\label{eq:MSCPT OBE13}
	\dot{\eta}_{13}(n_1, n_3)  	
	&=& 
	i\frac{\Omega_p^0}{2}\{ \sum_{n_1^{\prime\prime}} 
	\eta_{11}(n_1, n_1^{\prime\prime}) \mathcal{F}_{13}^{\,*} (n_1^{\prime\prime},n_3)
	- \sum_{n_3^{\prime\prime}} \mathcal{F}_{13}^{\,*} (n_1, n_3^{\prime\prime}) \eta_{33}(n_3^{\prime\prime}, n_3)\} \\ \nonumber
	&+& 
	i\frac{\Omega_q^0}{2} \sum_{n_2^{\prime\prime}} 
	\eta_{12}(n_1, n_2^{\prime\prime}) \mathcal{F}_{23}^{\,*} (n_2^{\prime\prime},n_3)
	+\left[i\{ (n_3+\Delta n)\nu_3 -n_1 \nu_1\}-\frac{R}{2}\right]\eta_{13}(n_1, n_3),\\
	\label{eq: MSCPT OBE23}
	\dot{\eta}_{23}(n_2, n_3)  	
	&=& 
	i\frac{\Omega_q^0}{2}\{ \sum_{n_2^{\prime\prime}} 
	\eta_{22}(n_2, n_2^{\prime\prime}) \mathcal{F}_{23}^{\,*} (n_2^{\prime\prime},n_3)
	- \sum_{n_3^{\prime\prime}} \mathcal{F}_{23}^{\,*} (n_2, n_3^{\prime\prime}) \eta_{33}(n_3^{\prime\prime}, n_3)\} \\ \nonumber
	&+&
	i\frac{\Omega_p^0}{2} \sum_{n_1^{\prime\prime}} 
	\eta_{21}(n_2, n_1^{\prime\prime}) \mathcal{F}_{13}^{\,*} (n_1^{\prime\prime},n_3)
	+\left[i\{ (n_3+\Delta n)\nu_3 -n_2 \nu_2+\delta_{\rm{CPT}}\}-\frac{R}{2}\right]\eta_{23}(n_2, n_3),\\
	\label{eq: MSCPT OBE12} 
	\dot{\eta}_{12}(n_1, n_2)
	&=&  	\frac{i}{2}\sum_{n_3^{\prime\prime}} 
	\left\{ \Omega_q^0 \eta_{13}(n_1, n_3^{\prime\prime}) \mathcal{F}_{32} (n_3^{\prime\prime},n_2)
	-  \Omega_p^0 \mathcal{F}_{13}^{\,*} (n_1, n_3^{\prime\prime}) \eta_{32}(n_3^{\prime\prime}, n_2)\right\}  \\
	&+&
	\{ i (n_2\nu_2-n_1\nu_1-\delta_{\rm{CPT}})-\gamma_{12} \} \eta_{12}(n_1, n_2),	\nonumber 
	\end{eqnarray}
\end{subequations}
$\dot{\eta}_{31}(n_3, n_1)  	=\dot{\eta}_{13}^{\,*} (n_1, n_3), 
\dot{\eta}_{32}(n_3, n_2)  	=\dot{\eta}_{23}^{\,*} (n_2, n_3)$,
and $\dot{\eta}_{21}(n_2, n_1)  	=\dot{\eta}_{12}^{\,*} (n_1, n_2)$.

\section{Recursion relations for ${\mathcal F} (n, l)$}
$\mathcal{F}$ factor in 1D is defined as
 \begin{equation}
{\mathcal F}_{ij} (n_i, n_j) 
= \langle \chi_i(n_i)| e^{i k_p x} |\chi_j(n_j) \rangle.
	\label{eq:FCF_DEF}
\end{equation}
As a specific example, we consider ${\mathcal F}_{31} (n_3, n_1)$, 
and  put $n_3 = n, n_1 =l$ and $k_p = k$, and omit the subscript $31$ from $\mathcal{F}$ for simplicity. 
Explicitly,
 \begin{equation}
	{\mathcal F} (n, l) 
	= C_{n} (a_3) C_l(a_1)
	\int_{-\infty}^{+\infty} H_{n}(a_3 x)H_{l} (a_1 x) e^{-\langle a^2 \rangle x^2+i k x }dx,
	\label{eq:FCF_A1}
\end{equation}
where $a_1 = \sqrt{{\nu_1 m}/{\hbar}\,}$ and $C_l(a_1) = \sqrt{{a_1}/{\pi^{1/2} \,2^{l} l! }\,}$ 
with $m$ being the atomic mass. 
$a_3$ and $C_n(a_3)$ are similarly defined, and $\langle a^2 \rangle = (a_1^2 +a_3^2)/2$.
Integrating by parts and using the recursion relations of the Hermite polynomials, we obtain the recursion relation for ${\mathcal F} (n, l) $,
\begin{equation}
	{\mathcal F} (n, l)
	=i \frac{ k a_1}{\langle a^2 \rangle}\sqrt{\frac{1}{2n}}{\mathcal F} (n-1, l)
	+\frac{\Delta a^2}{2\langle a^2 \rangle }\sqrt{\frac{n-1}{n}} {\mathcal F} (n-2, l)
	+\frac{a_1 a_3} {\langle a^2 \rangle} \sqrt{\frac{l}{n}} {\mathcal F} (n-1, l-1),
	\label{eq:recurson1}
\end{equation}
where $\Delta a^2 = a_3^2-a_1^2$. 
Alternatively,
\begin{equation}
	{\mathcal F} (n, l)
	=i \frac{k a_3}{\langle a^2 \rangle}\sqrt{\frac{1}{2l}}{\mathcal F} (n, l-1)
	-\frac{\Delta a^2}{2\langle a^2 \rangle }\sqrt{\frac{l-1}{l}} {\mathcal F} (n, l-2)
	+\frac{a_1 a_3} {\langle a^2 \rangle} \sqrt{\frac{n}{l}} {\mathcal F} (n-1, l-1).
	\label{eq:recurson2}
\end{equation}
When $\nu_1 = \nu_3$, $\Delta a^2 =0$ and the relation simplifies. 

To obtain ${\mathcal F}$ factor for a transition driven by a radio-frequency field, 
we substitute $k =0$ in Eq. (\ref{eq:FCF_A1}).
When $\nu_1 = \nu_3$, $\mathcal{F}(n, l ) = \delta_{nl}$. 
When $\nu_1 \neq \nu_3$, the recursion relations cannot be obtained by simply substituting $k =0$ in Eqs. (\ref{eq:recurson1}) and (\ref{eq:recurson2}), and a separate calculation yields
\begin{eqnarray}
	{\mathcal F} (n, l)
	&=&\sqrt{\frac{(n-1)(l-1)}{nl}}{\mathcal F} (n-2, l-2) +\frac{a_1 a_3} {\langle a^2 \rangle} \sqrt{\frac{1}{nl}} {\mathcal F} (n-1, l-1) \label{eq:recursion RF} \\
	&+&\frac{\Delta a^2}{2\langle a^2 \rangle }\sqrt{\frac{n-1}{n}} {\mathcal F} (n-2, l)
	-\frac{\Delta a^2}{2\langle a^2 \rangle }\sqrt{\frac{l-1}{l}} {\mathcal F} (n, l-2).
	\nonumber
\end{eqnarray}
In our previous publication on rf spectroscopy \cite{RF spectroscopy}, we used a series expansion of $H_n(a_i x)$  in terms of $(a_i-\bar{a})/\bar{a}$ with $\bar{a}=\sqrt{\langle a^2 \rangle}$  for $ i = 1,3$
to speed up the evaluation of ${\mathcal F} (n, l)$.
The recursion relation, Eq. (\ref{eq:recursion RF}), is much more efficient.


\newpage
\begin{thebibliography}{99}
\bibitem{Wineland1989}
F. Diedrich, J. C. Bergquist, W. M. Itano, and D. J. Wineland,
Laser Cooling to the Zero-Point Energy of Motion,
Phys. Rev. Lett. {\bf 62}, 403 (1989).	

\bibitem{VSCPT 1988}	
A. Aspect, E. Arimondo, R. Kaiser, N. Vansteenkiste, and C. Cohen-Tannoudji, 
Laser Cooling below the One-Photon Recoil Energy by Velocity-Selective Coherent Population Trapping,
Phys. Rev. Lett. {\bf 61}, 826 (1988).

\bibitem{gray molasses cooling}
G. Grynberg and J.-Y. Courtois, 
Proposal for a Magneto-Optical Lattice for Trapping Atoms in Nearly-Dark States,
Europhys. Lett. {\bf 27}, 41 (1994).

\bibitem{Levy flight}
F. Bardou, J. P. Bouchaud, O. Emile, A. Aspect, and C. Cohen-Tannoudji,
Subrecoil Laser Cooling and L{\'e}vy Flights,
Phys. Rev. Lett. {\bf 72}, 203 (1994).

\bibitem{M1 CPT}	
H. Kim, H. S. Han, T. H. Yoon, and D. Cho,
Coherent Population Trapping in a $\Lambda$ Configuration Coupled by Magnetic Dipole Interactions,
Phys. Rev. A {\bf 89}, 032507 (2014).

\bibitem{collisional dephasing}
Y. Sagi, I. Almog, and N. Davidson,
Universal Scaling of Collisional Spectral Narrowing in an Ensemble of Cold Atoms,
Phys. Rev. Lett. {\bf 105}, 093001 (2010).

\bibitem{MSCPT experiment}
S. Park, M. H. Seo, R. A. Kim, and D. Cho,
Motion-selective coherent population trapping by Raman sideband cooling along two paths in a $\Lambda$ configuration,	arXiv:2205.05224 [physics.atom-ph].

\bibitem{polar molecule}
L. Anderegg, B. L. Augenbraun, Y. Bao, S. Burchesky, L. W. Cheuk, W. Ketterle, and J. M. Doyle,
Laser Cooling of Optically Trapped Molecules,
Nat. Phys. {\bf 14} 890 (2018).

\bibitem{master eq for inverted Y}
J. Qi, 
Electromagnetically Induced Transparency in an Inverted Y-type Four-Level System,
Phys. Scr. {\bf 81}, 015402 (2010).

\bibitem{analogous Zeeman}	
D. Cho,
Analogous Zeeman Effect from the Tensor Polarizability in Alkali Atoms,
J. Korean Phys. Soc. {\bf 30}, 373, (1997).

\bibitem{Cohen Tanouudij}
C. Cohen-Tannoudji, J. Dupont-Roc, and G. Grynberg, 
{\it{Atom-Photon Interactions: Basic Processes and Applications}}
(John Wiley \& Sons, Inc., New York, 1992), pp. 518-524. 

\bibitem{parametric heating}
T. A. Savard, K. M. O'Hara, and J. E. Thomas,
Laser-noise-induced heating in far-off resonance optical traps,
Phys. Rev. A {\bf 56}, R1095 (1997).

\bibitem{RF spectroscopy}
S. Park, M. H. Seo, and D. Cho,
Ground-state hyperfine spectroscopy of $^{87}$Rb atoms in a 1D optical lattice,
J. Phys. B: At. Mol. Opt. Phys. {\bf{52}}, 235002 (2019).

\bibitem{dephasing_photon_scattering}
H. Uys, M. J. Biercuk, A. P. VanDevender, C. Ospelkaus, D. Meiser, R. Ozeri, and J. J. Bollinger,
Decoherence due to Elastic Rayleigh Scattering,
Phys. Rev. Lett. {\bf 105}, 200401 (2010).

\bibitem{power_servo}
F. Tricot, D. H. Phung, M. Lours, S.  Gu\'{e}randel, and E. de Clercq,
Power stabilizatinof a diode laser with an acousto-optic modulator,
Rev. Sci. Instrum. {\bf 89}, 113112 (2018).

\bibitem{cooling by anharmonicity}
N. Poli, R. J. Brecha, G. Roati, and G. Modugno,
Cooling atoms in an optical trap by selective parametric excitation,
Phys. Rev. A {\bf 65}, 021401(R), (2002).

\end{thebibliography}
\end{document}